\documentclass[11pt]{article}
\usepackage{amsmath,amssymb,bm, graphics,epsfig,subfigure,color}
\usepackage{amsthm,bbm,algorithm,algorithmic,url}

\usepackage{booktabs}

\newtheorem*{Lem*}{Lemma}

\numberwithin{equation}{section} \numberwithin{The}{section}

\newcommand{\Frac}[2]{\frac{\displaystyle#1}{\displaystyle#2}}

\newcommand{\ignore}[1]{}

\begin{document}
\baselineskip=16pt
\title{\bf Modeling Multivariate Cyber Risks: Deep Learning Dating Extreme Value Theory
}

\author{Mingyue Zhang Wu${^\star}$ ~~ Jinzhu Luo$^\diamond $ ~~ Xing Fang$^\diamond $ ~~ Maochao Xu$^{\star\star}$\thanks{Correspondence: {\tt mxu2@ilstu.edu}} ~~Peng Zhao${^\star}$  \\
~\\
$^\star$   School of Mathematics and Statistics, Jiangsu Normal University, China.\\
$^\diamond $ School of Information Technology,  Illinois State University, USA.\\
     $^{\star\star}$ Department of Mathematics, Illinois State University, USA\\
\date{\today}
}

\date{\today}
\maketitle

\begin{abstract}
Modeling cyber risks has been an important but challenging task in the domain of cyber security. It is mainly because of the high dimensionality and heavy tails of risk patterns. Those obstacles have hindered the development of statistical modeling of the multivariate cyber risks. In this work, we propose a novel approach for modeling the multivariate cyber risks which relies on the deep learning and extreme value theory. The proposed model not only enjoys the
high accurate point predictions via deep learning but also can provide the satisfactory high quantile prediction via extreme value theory.
The simulation study shows that the proposed model can model the multivariate cybe risks very well and provide satisfactory prediction performances.   The empirical evidence based on real honeypot attack data also shows  that the proposed  model has   very satisfactory prediction performances.

\end{abstract}


\section{Motivation and Introduction}
Cyber risk has become one of the most emerging risks in recent decades which can cause disastrous consequences and tremendous monetary losses \cite{jang2014survey,peng2018modeling,xu2019cybersecurity}. This is fundamentally caused by the fact that cyberspace is extremely difficult to secure because of its complexity
(e.g., an attacker can launch cyber attacks from anywhere in the world, and cyber systems have a very large ``surface'' of software vulnerabilities that can be exploited to penetrate into them).
What makes things worse is that cyber systems have become underlying pillars for various infrastructures and physical systems. For example, there are growing concerns about cyber threats to critical infrastructures because of the linkage between cyber and physical systems \cite{DHS2015}. In addition, cyber attacks causes  billions of dollars in losses each year based on the research by the Ponemon Institute and published by IBM Security \cite{ponemon2020}. In light of the risk and potential consequences of cyber risks, securing cyberspace has become not only a mission of homeland security but also an emerging task for private companies. This clearly calls for research into  cyber risk modeling, of which risk prediction is an important task.

In this work, we investigate a particular challenge that is encountered when modeling and analyzing cyber risk, namely the modeling of multivariate cyber risks. This
challenge is imposed by the high-dimensional dependence among cyber attacks and extreme attacks during a short time period. The dependence among cyber attacks is very common which is the nature of cyber risks. For example, the Distributed Denial of Service (DDoS)  attack, a large-scale DoS attack where the perpetrator often uses  thousands of hosts infected with malware to launch attacks to a target or multi targets (e.g., ports, computers, severs), is in nature high-dimensional and dependent. In particular, the DDos attacks are often extreme during a short time period. The defender often needs to estimate or predict the cyber risks for the purpose of adjusting the defense posture in practice. However, this is very challenging because of the high-dimensionality and heavy tails (i.e., extreme attacks during a short period) exhibited by cyber attack data \cite{zhan2013characterizing,zhan2015predicting,xu2017vine,peng2018modeling}, which hinders the development of modeling multivariate cyber risks. Therefore, in the literature of statistical modelings,  there are only few studies investigating multivariate cyber risks. For example, Xu et al. \cite{xu2017vine} proposed a vine copula approach for modeling the dependence among the time series of the number of cyber attacks, and the dependence between the time series of the number of attacked computers. Peng et al. \cite{peng2018modeling} developed a copula-GARCH model, which also uses vine copulas to model the multivariate dependence among cyber attacks. Ling et al. \cite{ling2019predicting}   proposed a vector autoregression (VAR) approach  to identifying a geospatial and temporal patterns in the cyberattacks by considering the long range dependence.

In recent years, there is a growing interest in developing deep learning models for time series forecasting \cite{langkvist2014review,deng2014deep,sezer2020financial}. For multivariate time series forecasting, Goel et al. \cite{goel2017r2n2} proposed  a hybrid R2N2 (residual recurrent neural networks) model, which
first models the time series with a simple linear model (e.g., VAR) and then models its residual errors using recurrent neural networks (RNNs). Lai et al. \cite{lai2018modeling} developed a  long- and short-term time-series network (LSTNet)  to model the multivariate time series. The main idea is to  use the convolution neural network  and the recurrent neural network (RNN) to model short-term local dependence and  long-term trends for time series. Che et al. \cite{che2018recurrent} studied a  deep learning model based on gated recurrent unit  (GRU) for modeling  multivariate time series with missing data.  Wang et al. \cite{wang2020deeppipe} presented an end-to-end framework called deep prediction interval and point estimation, which  can simultaneously perform point estimation and uncertainty quantification for multivariate time series forecasting. The main approach is to model the loss function by penalizing the loss of point estimation and the loss of prediction interval based on the deep learning model of long short-term memory network. To the best of our knowledge, there is no deep learning framework for modeling the multivariate cyber attack data in the domain of cyber security.

In this work, we propose a novel framework which is different from those in the literature
for modeling and predicting the  multivariate cyber attack data. Specifically, we develop a hybrid model by combining deep learning and extreme value theory \cite{de2007extreme,mcneil2010quantitative}. That is, we first model the multivariate attack time series via a deep learning model which aims to capture the high-dimensional dependence via the deep learning network. We then model the residuals exhibiting heavy tail via the extreme value theory. The proposed approach not only can provide accurate point prediction but also capture the tails very well (i.e., predicting the extreme attacks). This novel framework is particularly useful in practice as it can guide the defender to prepare the resource for both the regular attack (in terms of mean prediction) and worst attack (in terms of tail prediction) scenarios.

The rest of the paper is organized as follows. In Section 2, we present the preliminaries on the deep learning model and extreme value theory. In Section 3, we discuss the proposed the framework for model fitting and prediction. Section 4 assesses the proposed approach via simulation studies. In Section 5, we study two real honeypot attack data. In the final section, we conclude our results and present a discussion towards some limitations and future work.

\section{Preliminaries} \label{sec: preliminary}

\subsection{Long short-term memory (LSTM)}
The LSTM network, a special kind of RNNs,  introduced in \cite{hochreiter1997long} has
been received remarkable attention because of
the ability to learn long term patterns in sequential data and tremendously improve the prediction accuracy compared to other deep learning models. Therefore, this deep learning model has been successfully applied in many areas \cite{Goodfellow-et-al-2016}.
\begin{figure}[htbp!]
 \centering
 \includegraphics[width=.8\textwidth]{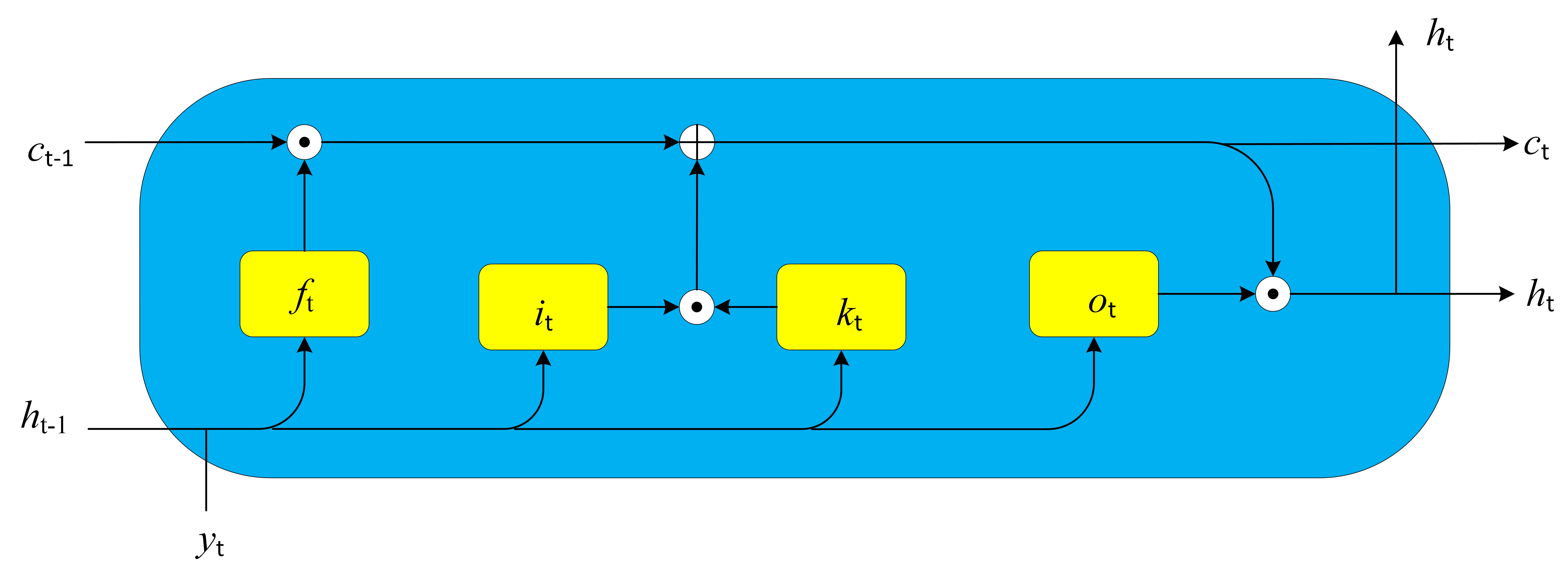}
  \caption{A LSTM block at step $t$.}
  \label{fig:lstm}
\end{figure}

 Figure \ref{fig:lstm} illustrates a LSTM block. Compared to regular RNN, the LSTM employs a different approach for the activation. Specifically, given time series data ${\bm y}_1,\ldots, {\bm y}_t$  where ${\bm y}_i=(y_{1,i},\ldots,y_{n,i})^\top$, $i=1,\ldots, t$, the activation $h_{t}$ of LSTM at step $t$ is computed based on four pieces: gate input (a.k.a, information gate), the forget gate, the output gate, and the cell gate. The information gate input at step $t$ is
\begin{equation*}
  i_{t} = \sigma(\mathbf{W}_{ia}\cdot h_{t-1}+\mathbf{W}_{iy}\cdot {\bm y}_t),
\end{equation*}
where $\sigma$ is  the sigmoid function, and $\mathbf{W}$'s are  the parameter matrices of the network learned during the training. The forget gate input and the output gate input are computed as
\begin{equation*}
  f_{t} = \sigma(\mathbf{W}_{fa}\cdot h_{t-1}+\mathbf{W}_{fy}\cdot {\bm y}_t),
\end{equation*}
\begin{equation*}
  o_{t} = \sigma(\mathbf{W}_{oa}\cdot h_{t-1}+\mathbf{W}_{oy}\cdot {\bm y}_t).
\end{equation*}

The cell gate input is computed as
\begin{equation*}
  C_{t} = f_{t}\odot C_{t-1} + i_{t}\odot k_{t},
\end{equation*}
where $\odot$ represents the element-wise Hadamard product,  the   $C_{t-1}$ is the cell state information from the previous step, and $k_{t}$ is
\begin{equation*}
  k_{t} = \text{tanh}(\mathbf{W}_{ca}\cdot h_{t-1}+\mathbf{W}_{cy}\cdot {\bm y}_t)
\end{equation*}
where $\text{tanh}$ is the hyperbolic tangent function. Finally, the activation at step $t$ is computed as
\begin{equation*}
  h_{t} = o_{t} \odot \text{tanh}(C_{t}),
\end{equation*}
where $h(t)\in \mathcal{R}^n$ is the final cell output.
\subsection{GRU and mLSTM}
As the variants of LSTM, GRU \cite{cho2014learning} and Mulitplicative LSTM (mLSTM) \cite{krause2016multiplicative} have also received much attention recently due to their extraordinary prediction abilities.
The GRU  uses less trainable parameters than LSTM, by having only two gates, the reset gate ($r_{t}$) and the update gate ($z_{t}$). Its computational process is as follows
\begin{align*}
z_{t} &= \sigma(\mathbf{W}_{zy} \cdot{\bm y}_t + \mathbf{W}_{za} \cdot h_{t-1}), \\
r_{t} &= \sigma(\mathbf{W}_{ry} \cdot {\bm y}_t + \mathbf{W}_{ra} \cdot h_{t-1}), \\
\hat{h}_t &= \text{tanh}(\mathbf{W}_{hy} \cdot {\bm y}_t + \mathbf{W}_{ha} (r_{t} \odot h_{t-1})), \\
h_t &=  (1 - z_t) \odot h_{t-1} + z_t \odot \hat{h}_t.
\end{align*}
Compared to the LSTM, the mLSTM  simply replaces $h_{t-1}$ with an intermediate state, $m_{t}$, in order to calculate $i_{t}$, $f_{t}$, $o_{t}$, and $k_{t}$, where $m_{t}$ is
\begin{equation*}
  m_{t} = \mathbf{W}_{ma}\cdot h_{t-1} \odot \mathbf{W}_{my}\cdot {\bm y}_t.
\end{equation*}

For more and detailed discussions on the deep learning models, please refer to \cite{hochreiter1997long,Goodfellow-et-al-2016}

\section{Deep learning framework for multivariate cyber risks}\label{sec:framwork}

Let ${\bm y}_t=(y_{1,t},\ldots,y_{n,t})^\top$ be a vector of attack time series. In practice, the dimension of $n$ can be very large, and particularly, the dependence among cyber  risks can be very nonlinear. This motivates us to use the following approach for analyzing multivariate cyber risks, which utilizes
the advantage of  high accurate point estimates of deep learning and the advantage of EVT for modeling the high quantiles. Specifically, the proposed model is as follows,
$${\bm y}_t=f({\bm y}_{t-1},\ldots,{\bm y}_{t-p})+{\bm \epsilon}_t,$$
where ${\bm \epsilon}_t=(\epsilon_{1,t},\cdots,\epsilon_{n,t})^\top$ is an unobservable zero mean white noise vector process, and $f$ is a function by mapping the vector $({\bm y}_{t-1},\ldots,{\bm y}_{t-p})$ to the mean vector of ${\bm y}_t$. Note that a special case is the VAR,
$${\bm y}_t={\bm \mu}_0+A_1 {\bm y}_{t-1}+\ldots+A_p {\bm y}_{t-p}+ {\bm \epsilon}_t,$$
where $A_i$ are   coefficient matrices, $i=1,\ldots,p$, and ${\bm \mu}_0$ is an $n$-dimensional constant vector. We propose to model the mean of ${\bm y}_t$ by
$$\hat {\bm y}_t=\hat f({\bm y}_{t-1},\ldots,{\bm y}_{t-p})$$
via deep learning approach, while the tails of ${\bm y}_t$ are estimated by EVT approach.

The proposed framework involves the following three stages.

\subsection*{Stage 1: Capturing the multivariate dependence via deep learning}

In multivariate time series modeling and prediction, the most challenging part is to find a suitable $f$ for capturing the complex dependence pattern. Deep learning as an emerging tool for multivariate time series modeling and forecasting has an extraordinary forecasting performance \cite{fang2019performance,goel2017r2n2}. Therefore, we develop the following procedure for modeling the mean function $f$ via deep leaning.

Given time series ${\bm y}_1,\ldots,{\bm y}_{t-1}$, we define $p$ as the number of input data points in each of them, and  $p$ also indicates the number of time steps to look back (i.e., lags). The prediction process can be formally described as:
 $$\hat {\bm y}_t=\hat f({\bm Y}_{t-1,p}),$$
where ${\bm Y}_{t-1,p}=({\bm y}_{t-1},\ldots,{\bm y}_{t-p})\in R^{n\times p}$ is the input matrix.
We apply the following objective function to measure the error generated by the model,
\begin{equation}
  J = \frac{1}{(T-t+1)n} \sum_{i=t}^{T} ||{\bm y}_i-\hat{{\bm y}}_i||_2^{2} + \lambda \cdot ||\mathbf{W}||^{2}_{2},
\end{equation}
where $\lambda$ is the tuning parameter, $||\cdot||_{2}^{2}$ represents the squared $L_2$ norm,
and $\mathbf{W}$ represents all the training parameters in the learning process. For instance, if the network is LSTM, then $\mathbf{W} = \{\mathbf{W}_{za}, \mathbf{W}_{zy}, \mathbf{W}_{ra}, \mathbf{W}_{ry}, \mathbf{W}_{ha}, \mathbf{W}_{hy}\}$.    The optimization is defined as
$$\mathbf{W}^*=\arg\min_{\mathbf{W}} J,$$
which can be solved by using the gradient descent method \cite{kingma2014adam}. To select the best deep learning model, we perform a manual grid search by setting the hyper-parameters of the networks as follows:
\begin{itemize}
    \item Network models: $M_1$-LSTM, $M_2$-mLSTM, $M_3$-GRU;
    \item Number of layers: \{1, 2, 3\}; Size of each layer: \{16, 32, 64\};
    \item Batch size: \{5,10\};  Bidirectional: \{True, False\}; $p=\{1,2,3,4,5\}$;
    \item Penalty parameter: $\lambda = \{0.01, 0.001\}$; Initial learning rate: $\gamma = \{0.01, 0.001\} $;
    \item Number of training epochs: \{40, 50, 60, 70, 80\}.
\end{itemize}

Algorithm \ref{alg1} presents the detailed procedure for selecting the best deep learning model.


{\small \begin{algorithm}[!hbtp]                      
\caption{Algorithm for deep learning model training and selection.}          
\label{alg1}                           
\begin{flushleft}
    INPUT:  Historical time series data:   training part $\{(t,{\bm y}_{t})| t=1,\ldots,m\}$, and validation part $\{(t,{\bm y}_{t})|t=m+1,\ldots,T\}$; model $M = \{M_1, M_2, M_3\}$; layers $l=\{1, 2, 3\}$; size $s=\{16, 32, 64\}$; bidirectional $d= \{{\rm True}, {\rm False}\}$; epoch $b=\{40, 50, 60, 70, 80\}$;  $\lambda=\{0.01, 0.001\}$; $A = \phi$.
\begin{algorithmic}[1]                    
\FOR{$p\in\{1,2,3,4,5\}$, $r\in \{5, 10\}$}
    \STATE Split the data set into mini-batch of size $r$;
    \FOR{$\gamma \in \{0.01, 0.001\}$, $\lambda \in \{0.01, 0.001\}$}
        \FOR{$i \in \{1,2,3\}$, $d= \{{\rm True}, {\rm False}\}$}
            \FOR{$l \in \{1, 2, 3\}$, $s \in \{16, 32, 64\}$, $b\in\{40, 50, 60, 70, 80\}$}
                \STATE Randomly initialize $M_i$ based on $l$, $s$, and $d$, with all the parameters saved in         $\mathbf{W}$;
            \ENDFOR
            \STATE $j \leftarrow 0$;
            \WHILE {$j <=b$ }
            \FOR {each batch from the training data}
                \STATE Compute   $J$  by performing forward propagation using $\lambda$;
                       \STATE { Update $\mathbf{W}$ using gradient descent based on $J$ with the learning rate $\gamma$;}
            \ENDFOR
            \STATE Compute mean square error (MSE) by performing forward propagation on the validation data;
            \IF{MSE is not dropping}
                \STATE $A \leftarrow M_i$;
                \STATE break;
            \ENDIF
            \STATE $j \leftarrow j+1$;
            \ENDWHILE
        \ENDFOR
    \ENDFOR
\ENDFOR
\RETURN $M^* \in A$, where $M^*$ has the lowest MSE;
\end{algorithmic}
 OUTPUT: Deep learning model $M^*$ with the corresponding parameters  ${\bm W}^*$, and fitted values $\hat {\bm y}_{t}, t=1,\ldots,T$.
\end{flushleft}
\end{algorithm}
}



\subsection*{State 2: Modeling high quantiles via extreme value theory}
After fixing the deep learning model, the fitted values at time $t$ are
$$\hat {\bm y}_{t}=\hat f({\bm Y}_{t-1,p}),$$
and the residuals are
$${\bm e}_{t}= {\bm y}_{t}-\hat {\bm y}_{t},$$
where ${\bm e}_{t}=(e_{1,t},\ldots,e_{n,t})^\top$.
The second stage is to model the residuals by some statistical distribution. Since the heavy tails are often observed in the attack data, we propose to model
the high quantiles of the residuals by the EVT approach. This is in principle in line with the  two-stage pseudo-maximum-likelihood approach in \cite{mcneil2000}.

Recall that a popular EVT method is known as the {\em peaks over threshold} approach  \cite{mcneil2010quantitative,de2007extreme}. Specifically, given a sequence of i.i.d.
observations $X_1,\ldots,X_n$,
the excesses $X_i-\mu$ with respect to some suitably large threshold $\mu$ can be modeled by, under certain mild conditions, the {\em Generalized Pareto Distribution} (GPD). The  survival function of the GPD is:
\begin{equation}\label{eq:gpd}
\bar G(x)=1-  G(x)=\left\{\begin{array}{cc}
  \left(1+\xi \dfrac{x-\mu}{\sigma}\right)_+^{-1/\xi}, & \xi\ne 0, \\
  \exp\left\{-\Frac{x-\mu}{\sigma}\right\}, & \xi=0.
\end{array}\right.,
\end{equation}
where $x\ge \mu$ if $\xi\in\mathbb{R}^+$ and $x\in [\mu,\mu-\sigma/\xi]$ if $\xi\in \mathbb{R}^{-}$,  and $\xi$ and $\sigma$ are respectively called the {\em shape} and {\em scale} parameters.



\subsection*{Stage 3: Prediction and and evaluations}
After we determine the deep learning model and residual distribution, then we can use the forward propagation to predict the mean
$$\hat {\bm y}_{t+1}=\hat f({\bm Y}_{t,p}).$$
The $q$-quantile  of $y_{i,t+1}$ can be predicted as
 \begin{equation}\label{eq:var}
 \hat {\bm y}_{i,t+1,q}=\hat {y}_{i,t+1}+G_i^{-1}(q),
 \end{equation}
where $G_i^{-1}(q)$ is the $q$-quantile of $G_i$ as defined in Eq. \eqref{eq:gpd} for time series $i$, $i=1,\ldots,n$.  Algorithm \ref{alg2} presents the detailed procedure for the prediction.
{\small \begin{algorithm}[!hbtp]                      
\caption{Algorithm for predicting the mean and high quantiles.}          
\label{alg2}                           
\begin{flushleft}
    INPUT: Deep learning model $M^*$ from Algorithm \ref{alg1}; fitted values  $\{(t,\hat{\bm  y}_{t})|t=1,\ldots,T\}$; training and validation data $\{(t,{\bm y}_{t})|t=1,\ldots,T\}$; testing data
    $\{(t,{\bm y}_{t})|t=T+1,\ldots,N\}$; $q$.
\begin{algorithmic}[1]                    
    \FOR {$j=T\ldots,N-1$}
  \STATE Compute residuals ${\bm e}_t={\bm y}_{t}-\hat {\bm y}_{t}$, $t=1,\ldots,j$;
  \STATE Fit $G_i$ in Eq. \eqref{eq:gpd} to $e_{i,t}$ by setting threshold $\mu_i$, $i=1,\ldots,n$, $t=1,\ldots,j$;
        \STATE Predict   $\hat{{\bm y}}_{j+1}$  by performing forward propagation based on Model $M^*$;
        \STATE Predict the  $q$-quantile  of $\hat y_{i,j+1,q}$ by using Eq. \eqref{eq:var};
    \ENDFOR
    \RETURN  $\{(t,{\bm y}_{t})|t=T+1,\ldots,N\}$;
\end{algorithmic}
 OUTPUT: Predicted values $\hat {\bm y}_{t}$ and $q$-quantile $ \hat {\bm y}_{t,q}$, $t=T+1,\ldots,N$.
\end{flushleft}
\end{algorithm}
}

The prediction performance for the mean prediction are evaluated based on MSE and  mean absolute percentage error (MAPE) \cite{hyndman2006another}.

In order to assess the prediction accuracy of quantiles,  we propose using the Value-at-Risk (VaR) \cite{mcneil2010quantitative} metric because it is directly related to the quantities of interest.  Recall that for a random variable $X_t$, the VaR at level $\alpha$ for some $0<\alpha<1$ is defined as
${\rm VaR}_\alpha(t)=\inf\left\{l: P\left(X_t \leq l\right)\ge \alpha\right\}$.
An observed value that is greater than the predicted ${\rm VaR}_\alpha(t)$ is called a {\em violation}, indicating inaccurate prediction. In order to evaluate the prediction accuracy of the VaR values, we adopt the following two popular tests \cite{christoffersen1998evaluating}:
(i) the unconditional coverage test, denoted by LR$_{uc}$, which evaluates whether or not the fraction of violations is significantly different from the model's violations; (ii) the conditional coverage test, denoted by LR$_{cc}$, which is a joint likelihood ratio test for the independence of violations and unconditional coverage.

\section{Simulation study}\label{sec:simulation}
In this section, we perform a simulation study on assessing the performance of proposed approach. To mimicking the attack data with heavy tails, we randomly generate two data sets with size $5000\times 5$ from the following models.  In each experiment, the simulated data is split into three parts: the first 3500 is used for training, and the following  500 observations is used for the validation. The testing  data is the last 1000 observations.
\begin{itemize}
  \item VAR with heavy tail. The VAR model is set as follows.
  $${\bm y}_t={\bm \mu}_0+A_1 {\bm y}_{t-1}+ A_2 {\bm y}_{t-2}+ {\bm \epsilon}_t,$$
  where $t=1,\ldots,5000$, ${\bm \mu}_0=(\mu_{1,0},\ldots,\mu_{5,0})$, ${\bf y}_t=(y_{1,t},\ldots,y_{5,t})$, $A_1$ and $A_2$ are $5\times 5$ coefficient matrices. In the experiment,  $A_1$ and $A_2$ are generated from uniform distribution $[-0.2,0.2]$, and $\mu_{i,0}=100$,   $i=1,\ldots,5$. The error ${\bm \epsilon}_t=(\epsilon_{1,t},\ldots,\epsilon_{5,t})$ has zero mean, and $\epsilon_{i,t}/\sigma$, $i=1,\ldots,5$, follows a skewed-t distribution with density   as follows
  \begin{equation}\label{eq:sstd}
    g(z)=\frac{2}{\xi+\xi^{-1}}\left[t_\nu (\xi z){\rm I}(z<0)
+t_\nu v\left(\xi^{-1} z\right) {\rm I}(z\ge 0)\right],
  \end{equation}
where ${\rm I}(\cdot)$ is the indicator function,  $\xi>0$ is the skewness parameter, and
$$t_{\nu}(z)=\frac{\Gamma((\nu+1)/2)}{\sqrt{\nu\pi}\Gamma(\nu/2)}
\left[1+z^2/\nu\right]^{-(\nu+1)/2}$$
with  the shape parameter $\nu>0$. In the experiment, the parameters are set as $\sigma=20$,  $\xi=1.5$, and $\nu=3$.

Figure \ref{fig:var-ts} plots the simulated time series, and we observe there exist large values which indicate the heavy tail.
\begin{figure}[htbp!]
\centering
\subfigure[Time series 1]{\includegraphics[width=0.32\textwidth]{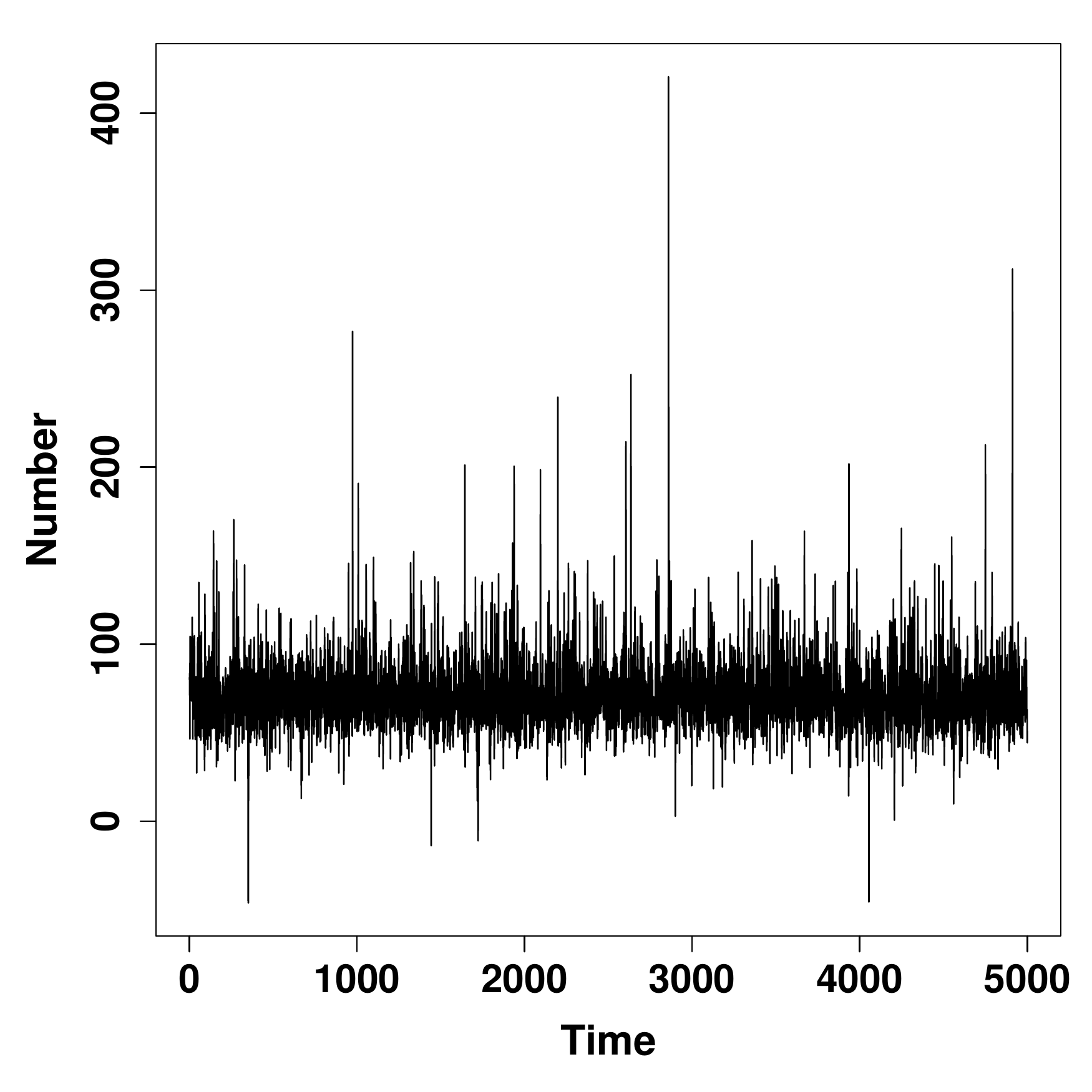}}
\subfigure[Time series 2]{\includegraphics[width=0.32\textwidth]{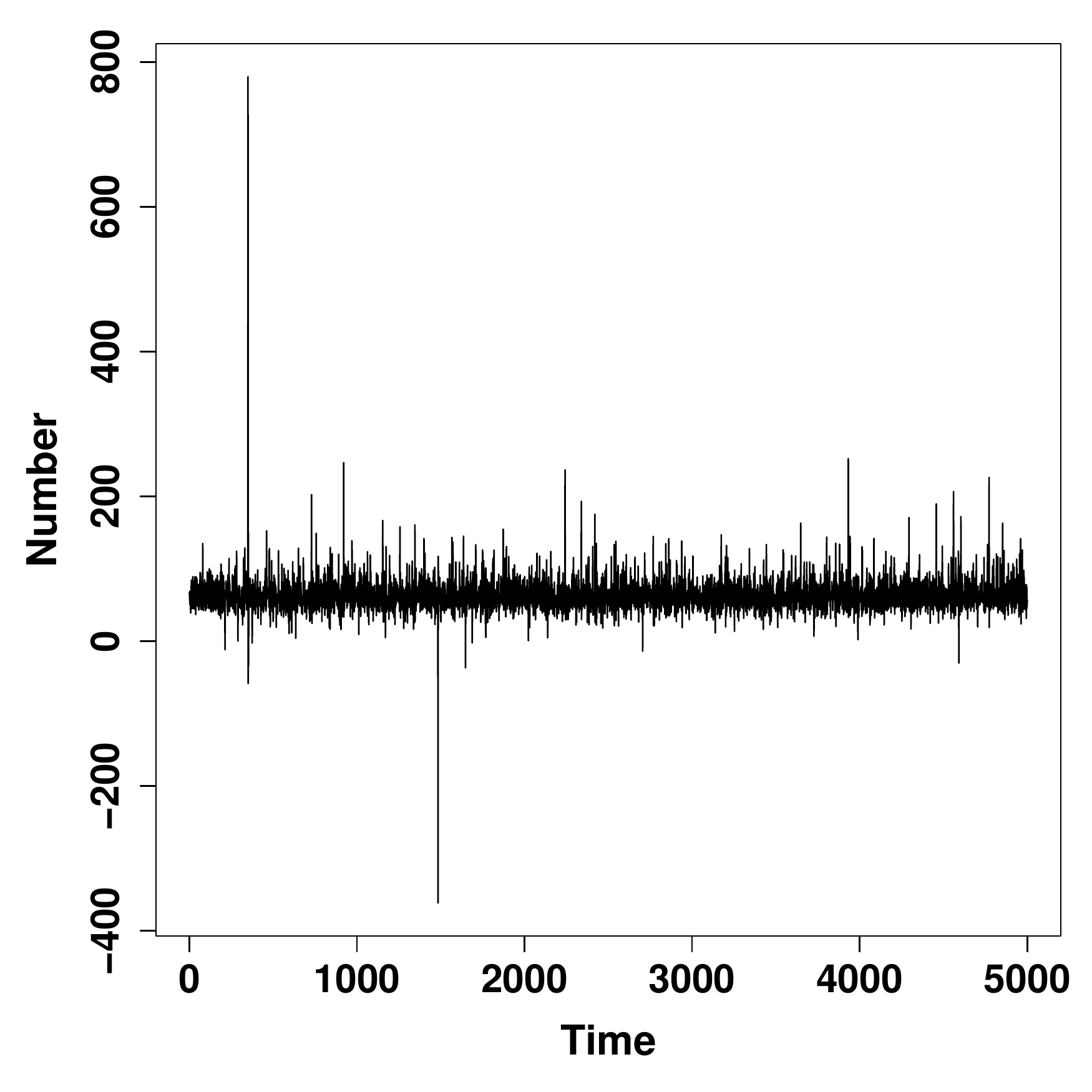}}
\subfigure[Time series 3]{\includegraphics[width=0.32\textwidth]{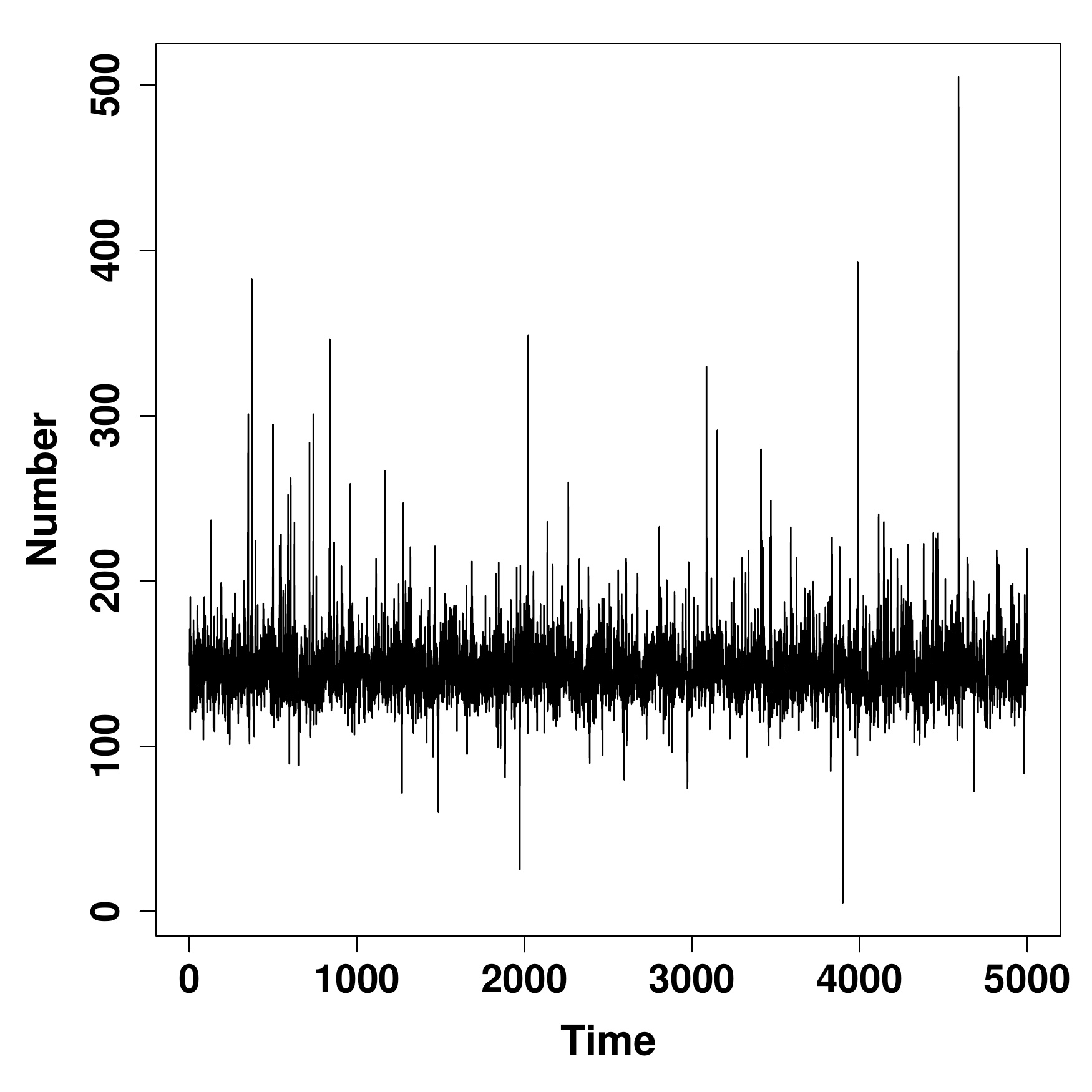}}
\subfigure[Time series 4]{\includegraphics[width=0.32\textwidth]{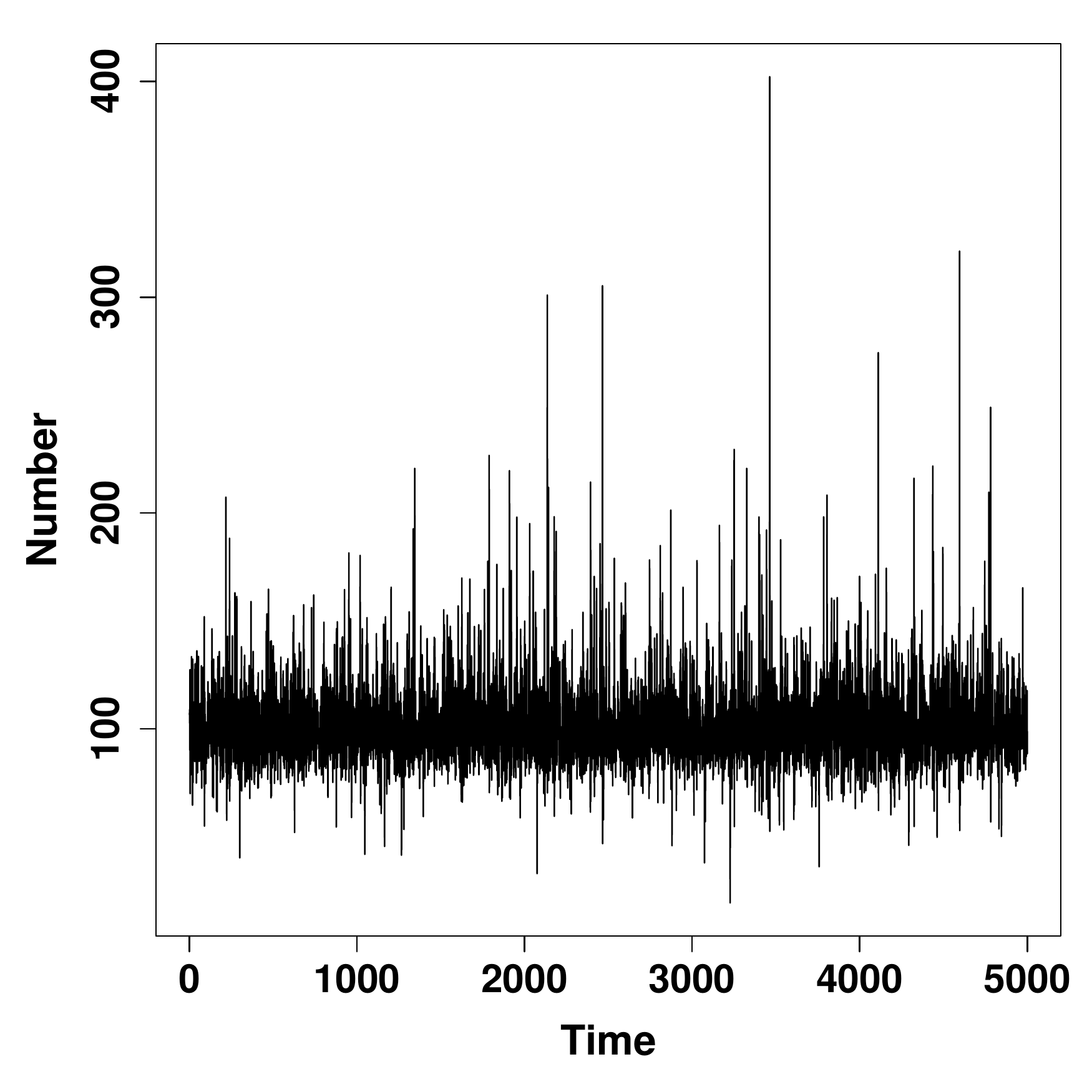}}
\subfigure[Time series 5]{\includegraphics[width=0.32\textwidth]{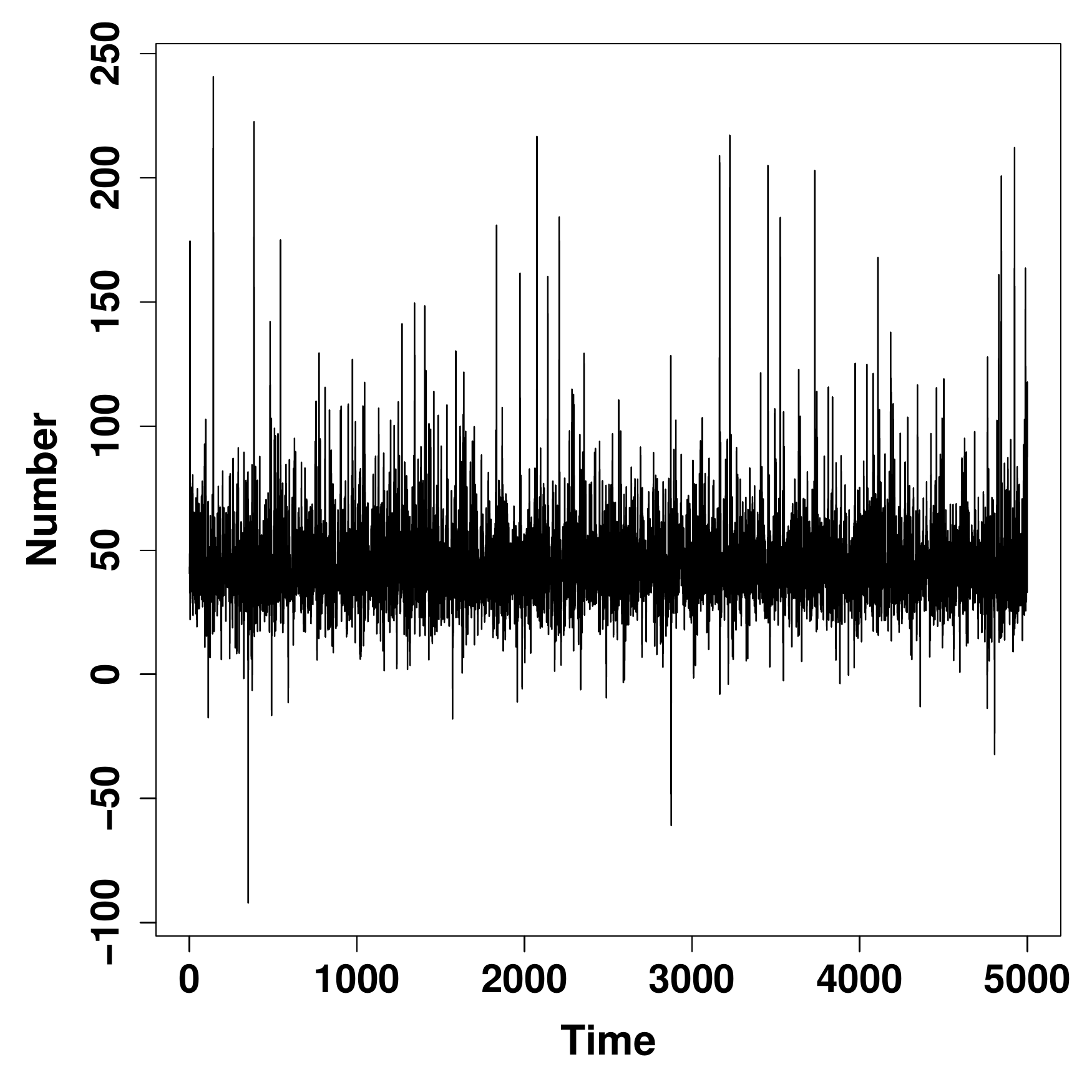}}
\caption{Time series plots of simulated VAR with skewed-t tails. \label{fig:var-ts}}
\end{figure}
We employ Algorithm \ref{alg1} to train and validate the deep learning model on the training and validation data sets. The residuals are used to fit the distribution $G$ in Eq. \eqref{eq:gpd} with the threshold setting to be $.9$-quantile of residuals for each time series which is determined via the mean residual life  plot \cite{mcneil2010quantitative}.

The prediction is performed via Algorithm \ref{alg2}, where the threshold is set to  $.9$-quantile of residuals. The prediction performance is reported in Table \ref{table:var-test1}. For the comparison purpose, we also report the prediction performance of VAR, where the lag $p$ is selected via AIC criterion. The VAR model is considered as a benchmark model in the sequel discussion.  It is seen from Table \ref{table:var-test1} that the proposed model outperforms the benchmark model for all the time series in terms of MAPE and MSE for the point prediction. Particularly, the MSEs are significantly improved by the deep learning approach.

\begin{table}[htbp!]\centering
\centering
\caption{Prediction performances based on MAPE and MSE for the proposed model and benchmark model.  \label{table:var-test1}}
\begin{tabular}{l|cc|cc}
  \toprule
 Series &\multicolumn{2}{|c|}{Deep}   &\multicolumn{2}{|c}{Benchmark} \\ \midrule
 &MAPE&MSE &MAPE&MSE \\\midrule
    1  & 0.1116 & 225.4169 & 0.1768 & 356.6563 \\ \midrule
    2  & 0.1152 & 211.2983 & 0.1908 & 357.2598 \\ \midrule
    3  & 0.0542 & 285.8806 & 0.0841 & 408.9762 \\ \midrule
    4  & 0.0771 & 226.0736 & 0.123 & 382.4075 \\ \midrule
    5  & 0.1716 & 191.3857 & 0.2839 & 378.5063 \\
 \bottomrule
\end{tabular}
\end{table}

\begin{figure}[htbp!]
\centering
\subfigure[Violations-Deep+EVT]{\includegraphics[width=0.32\textwidth]{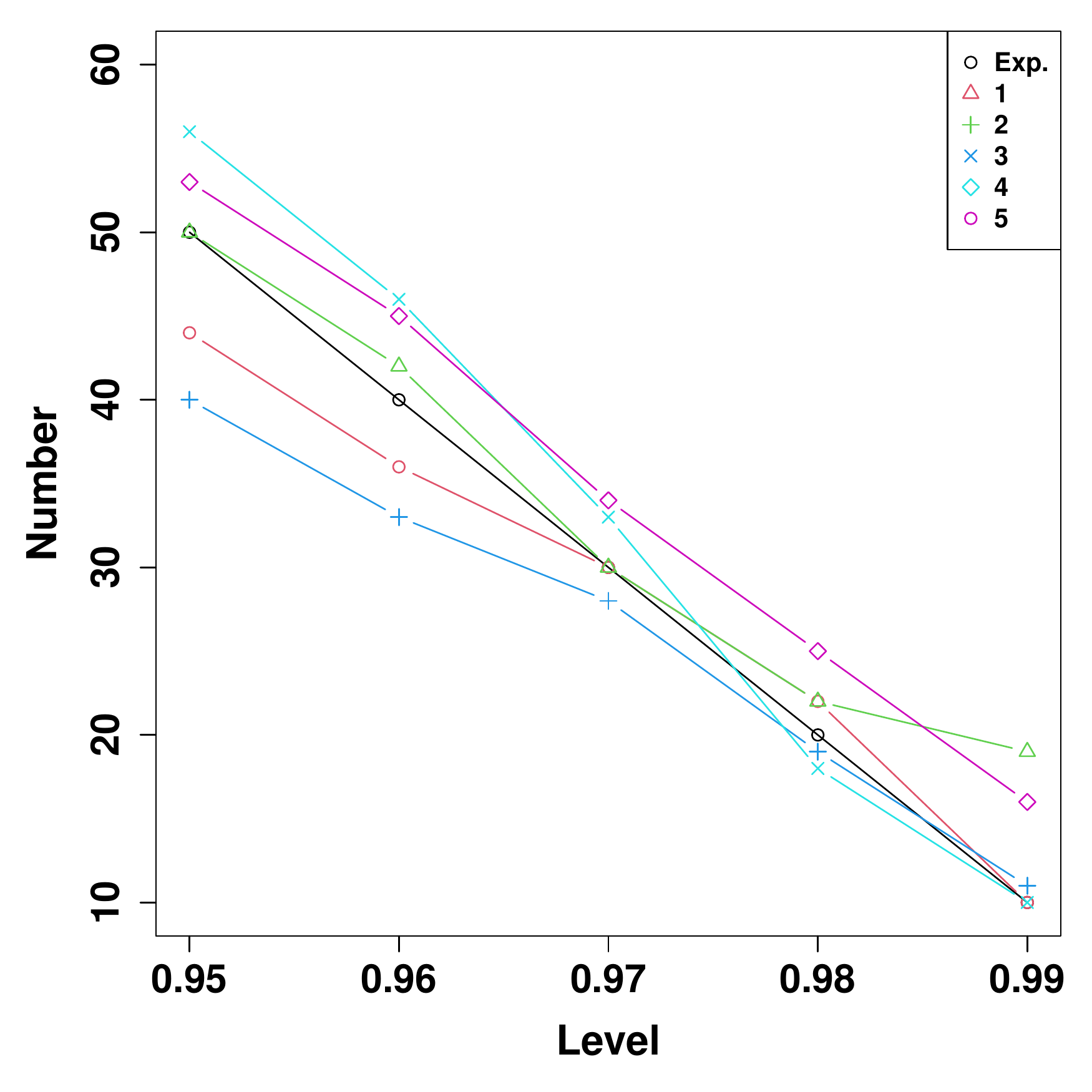}\label{fig:var-deep-vio}}
\subfigure[LRuc-Deep+EVT]{\includegraphics[width=0.32\textwidth]{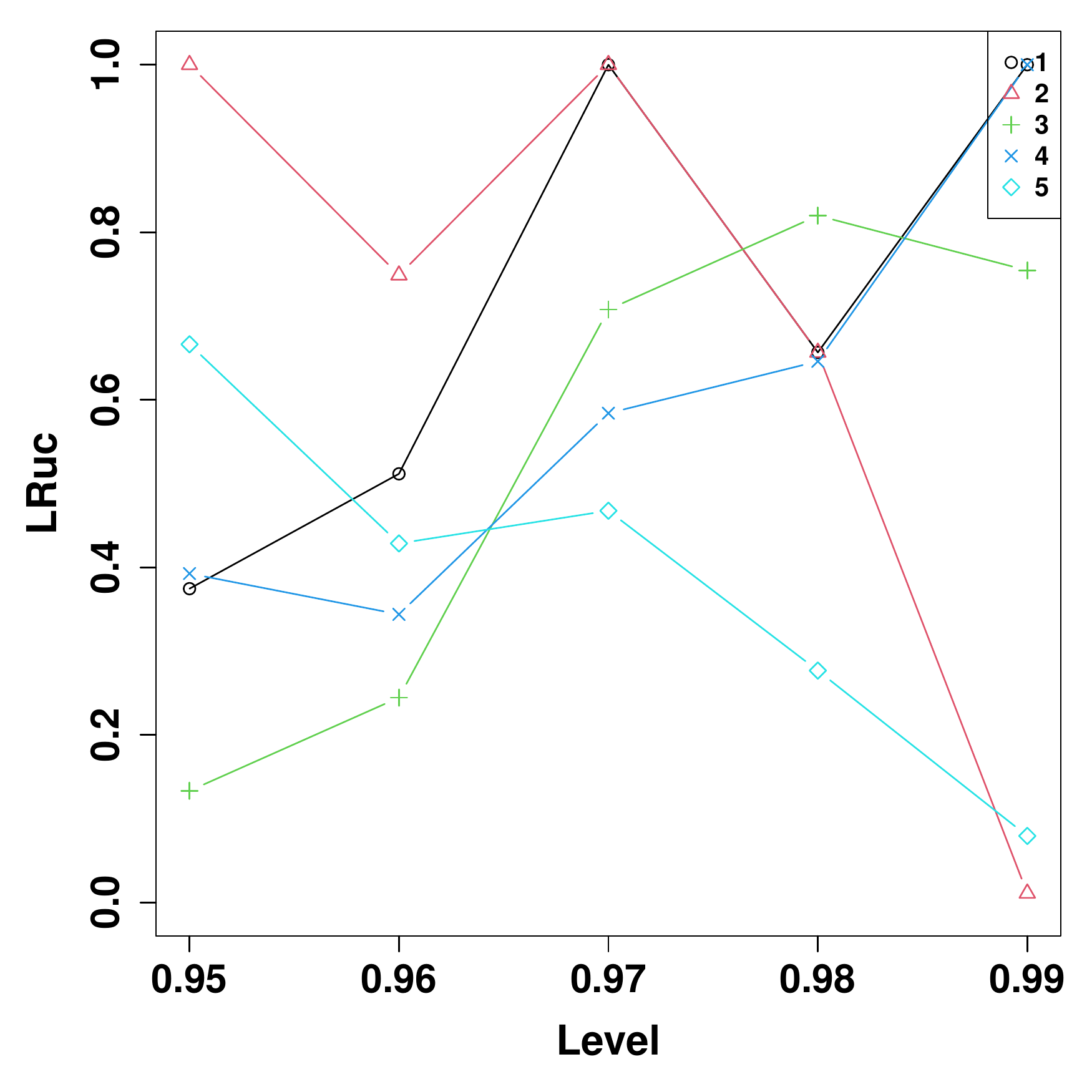}\label{fig:var-deep-lruc}}
\subfigure[LRcc-Deep+EVT]{\includegraphics[width=0.32\textwidth]{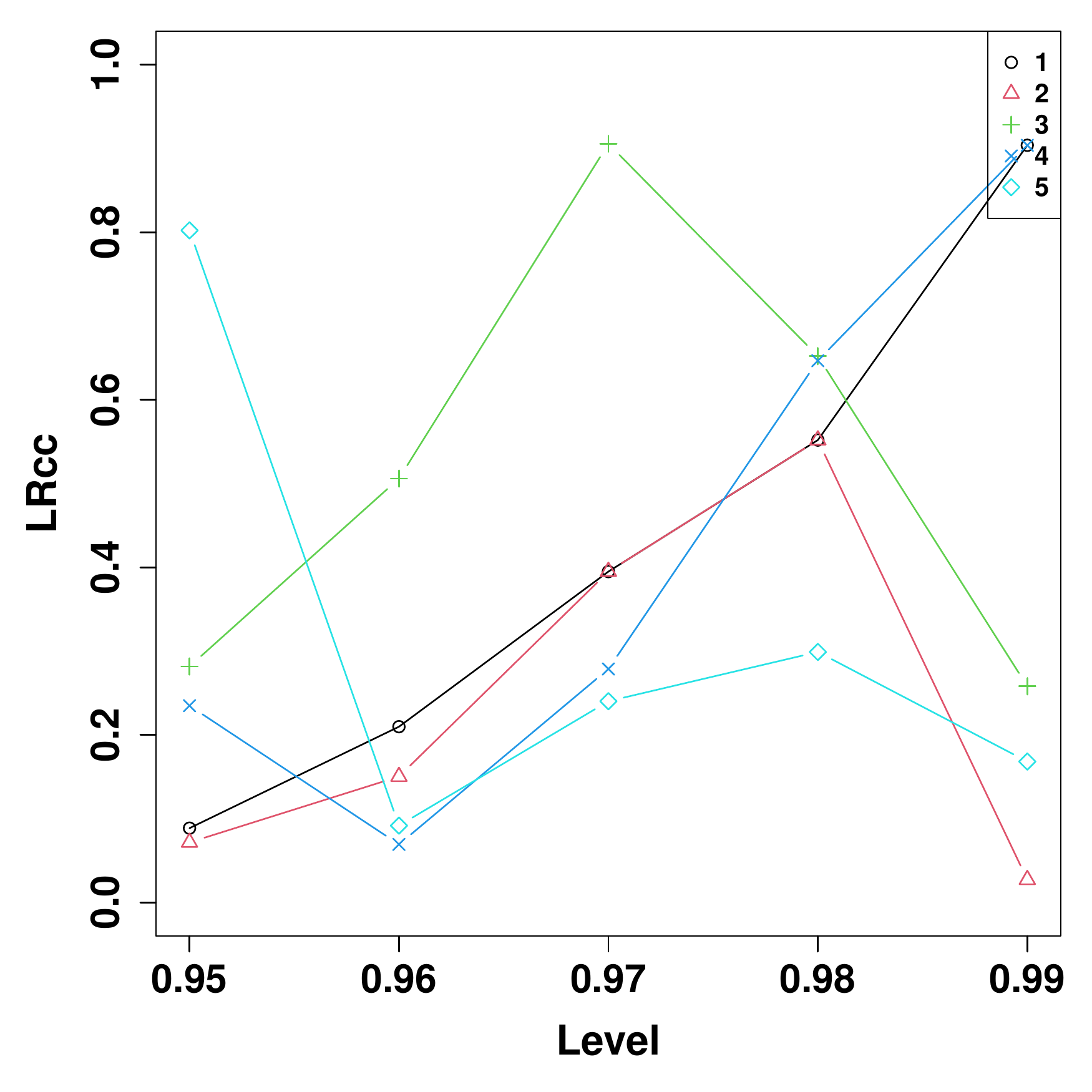}\label{fig:var-deep-lrcc}}
\subfigure[Violations-Benchmark]{\includegraphics[width=0.32\textwidth]{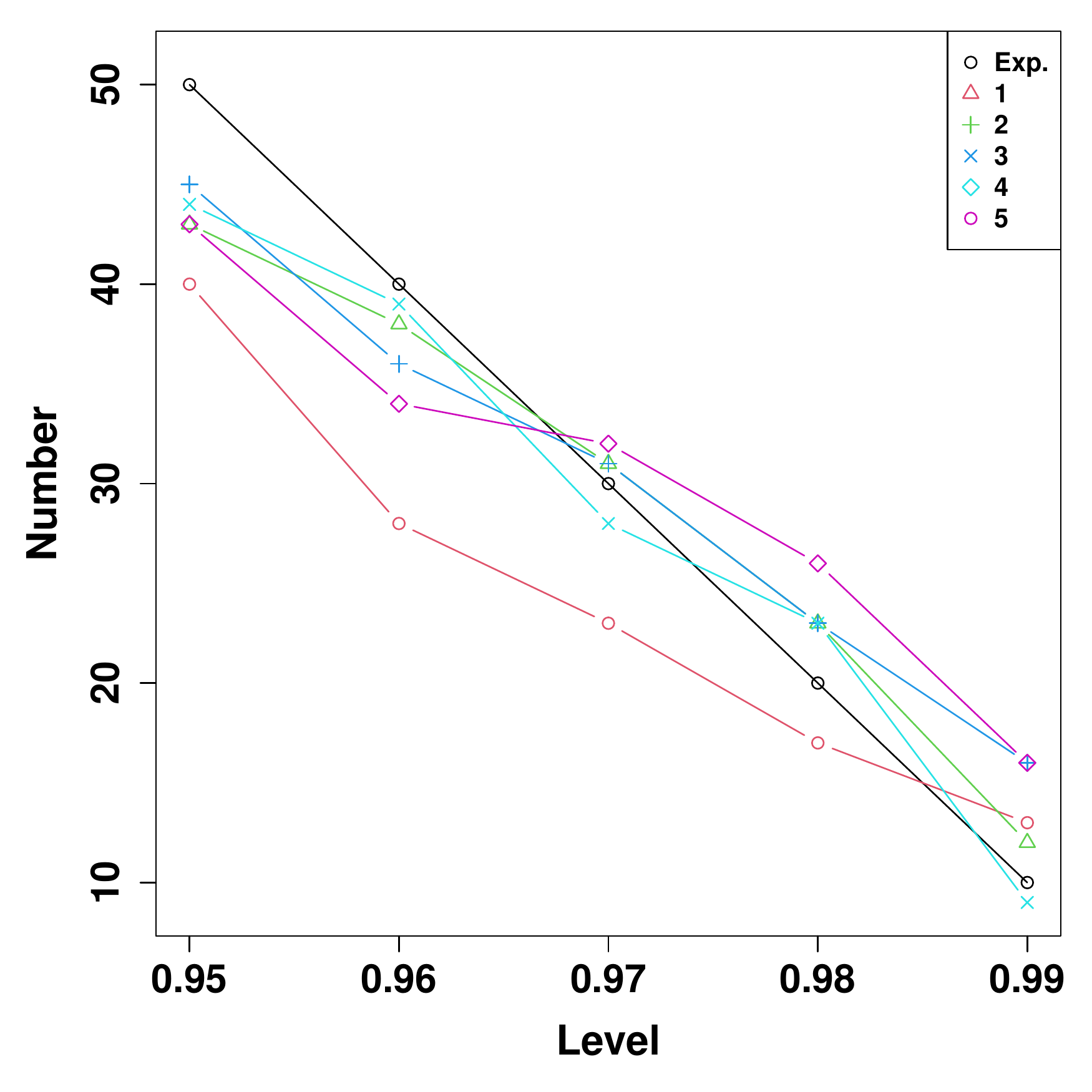}\label{fig:var-ben-vio}}
\subfigure[LRuc-Benchmark]{\includegraphics[width=0.32\textwidth]{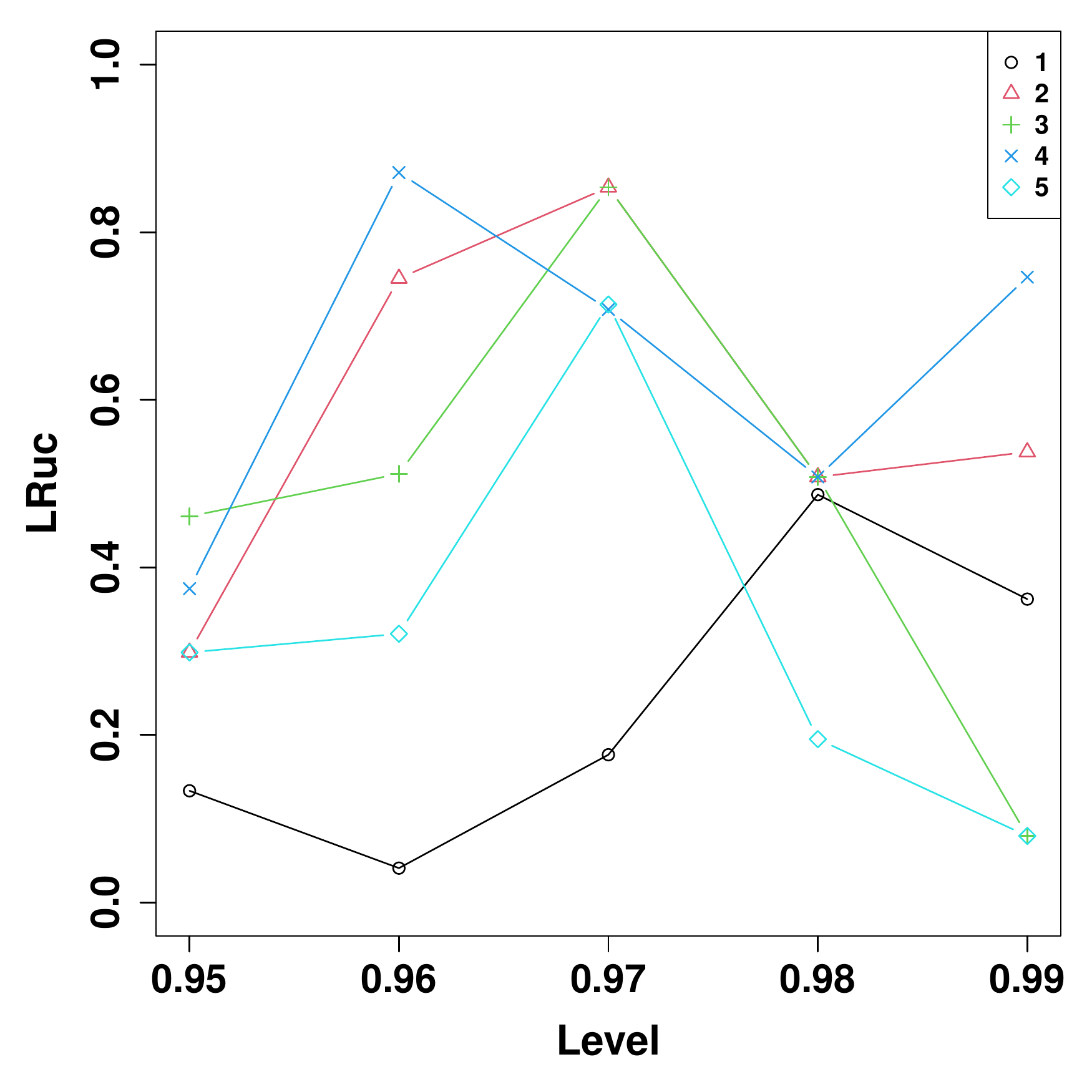}\label{fig:var-ben-lruc}}
\subfigure[LRcc-Benchmark]{\includegraphics[width=0.32\textwidth]{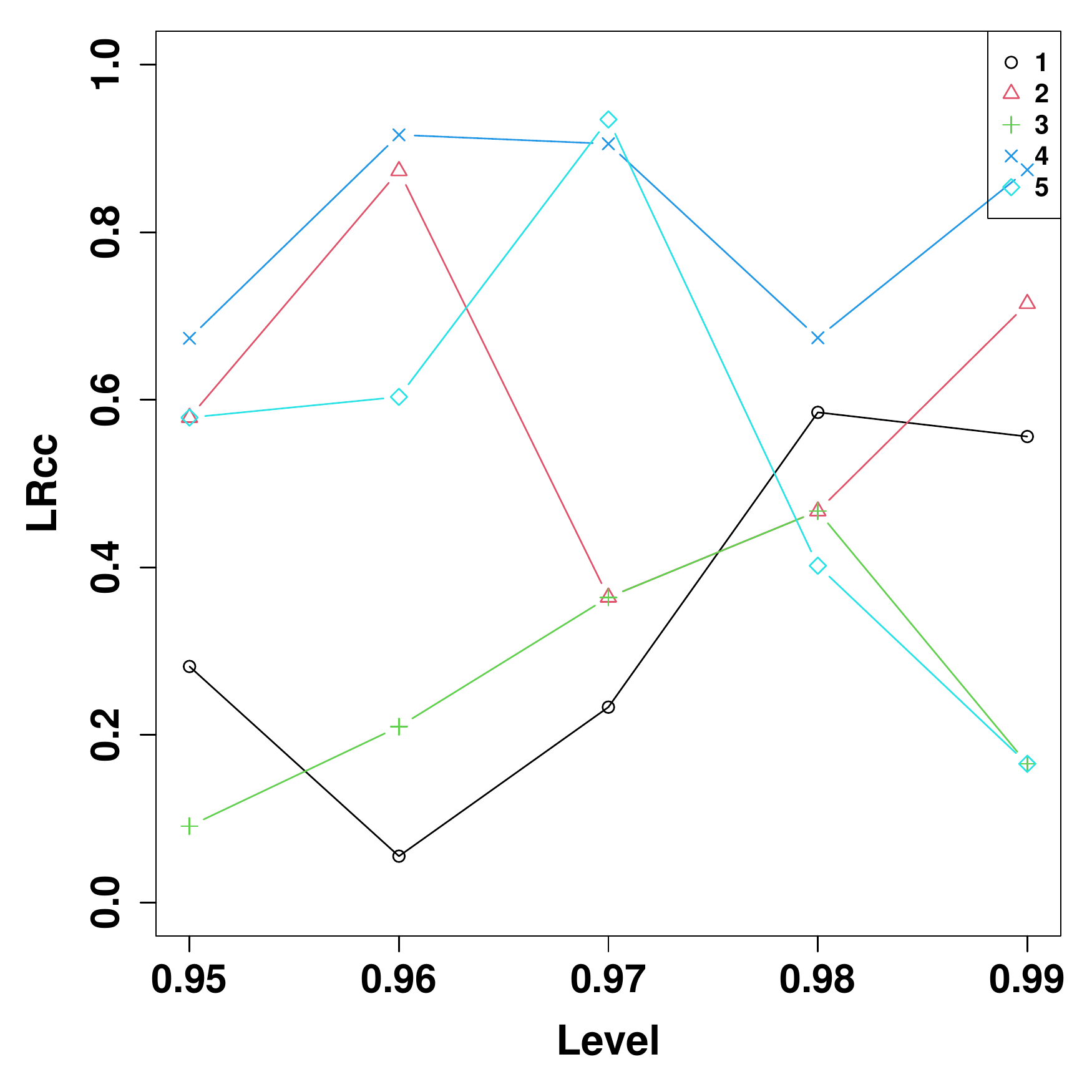}\label{fig:var-ben-lrcc}}
\caption{Numbers of violations and $p$-values of LRuc and LRcc tests for the proposed model and benchmark model, where 'Exp.' represents expected number of violations, and $x$-axis represents the VaR level $\alpha$. \label{fig:var-data1-violation}}
\end{figure}

For the quantiles prediction, we plot the violations for the proposed model and benchmark model in Figure \ref{fig:var-deep-vio} and \ref{fig:var-ben-vio}, respectively. It is seen that the observed number of violations are fairly close to the expected ones by the proposed approach from $.95$ to $.98$ levels. The benchmark model underestimates the number of violations at the $.95$ level. For the $.99$ level, it is seen that the proposed approach underestimates the number of violations for time series 4 and 5, which result in relatively smaller $p$-values of LRuc and LRcc tests in Figures \ref{fig:var-deep-lruc} and \ref{fig:var-deep-lrcc}. For the benchmark model, we observe that it underestimates four time series at level $.99$ from Figure \ref{fig:var-ben-vio}.

We conclude that the proposed approach has an overall satisfactory prediction performance for high quantiles from level $.95$ to $.98$.

  \item Copula+AR-GARCH with heavy tail.  For this simulation, we assume that the mean part follows AR(1) process
  \begin{equation*}\label{argarch}
  Y_{i,t}-\mu= \phi_{1}   (Y_{i,t-1}-\mu) +\epsilon_{i,t},
\end{equation*}
where $i=1,\ldots,5$, and
$$\epsilon_{i,t}=\sigma_{i,t} Z_{i,t}$$
with $Z_{i,t}$ being the innovations that are identically distributed with skewed-t density $g(\cdot)$ in Eq. \eqref{eq:sstd} and the dependence structure is specified via R-vine copula, and $\sigma_{t}$ follows a standard GARCH(1,1) model, i.e.,
$$\sigma_{i,t}^2=w+ \alpha_{1} \epsilon^2_{i,t-1}+  \beta_{1} \sigma^2_{i,t-1}.$$
 In the experiment, the parameters are set as follows
 $$(\mu,\phi_1,w,\alpha_1,\beta_1,\xi,\nu)=(50,.6,.5,.05,.8,1.5,3).$$
To simulate the dependence among $Z_{i,t}$, we generate 5-dimension multivariate uniform distribution, where the dependence structure is specified via the R-vine copula. Specifically, the R-vine tree matrix is as follows
$$ \begin{pmatrix}
2 & 0 & 0 & 0 & 0 \\
5 & 3 & 0 & 0 & 0 \\
3 & 5 & 4 & 0 & 0 \\
1 & 1 & 5 & 5 & 0 \\
4 & 4 & 1 & 1 & 1 \\
\end{pmatrix} $$
  and the family matrix is set as
  $$\begin{pmatrix}
0 & 0 & 0 & 0 & 0 \\
1 & 0 & 0 & 0 & 0 \\
3 & 3 & 0 & 0 & 0 \\
4 & 4 & 4 & 0 & 0 \\
4 & 1 & 1 & 3 & 0 \\
\end{pmatrix}.$$
The parameter matrix is set as follows
$$\begin{pmatrix}
0 & 0 & 0 & 0 & 0 \\
0.2 & 0 & 0 & 0 & 0 \\
0.9 & 1.1 & 0 & 0 & 0 \\
1.5 & 1.6 & 1.9 & 0 & 0 \\
3.9 & 0.9 & 0.5 & 4.8 & 0 \\
\end{pmatrix}.$$
A $5000\times 5$ matrix from the above R-vine structure is generated, which is used for generating the dependent $Z_{i,t}$,
$i=1,\ldots,5$.

\begin{figure}[htbp!]
\centering
\subfigure[Time series 1]{\includegraphics[width=0.32\textwidth]{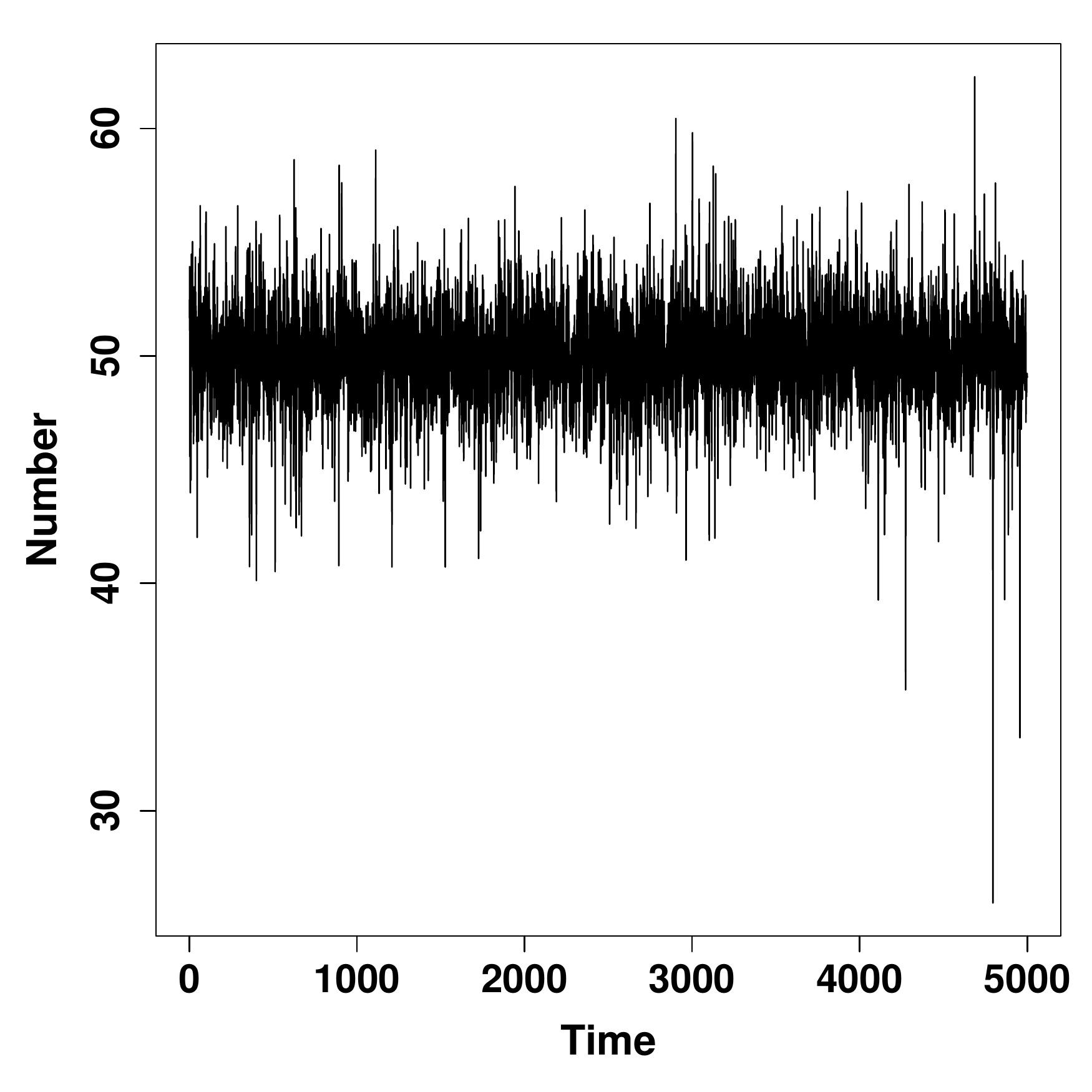}}
\subfigure[Time series 2]{\includegraphics[width=0.32\textwidth]{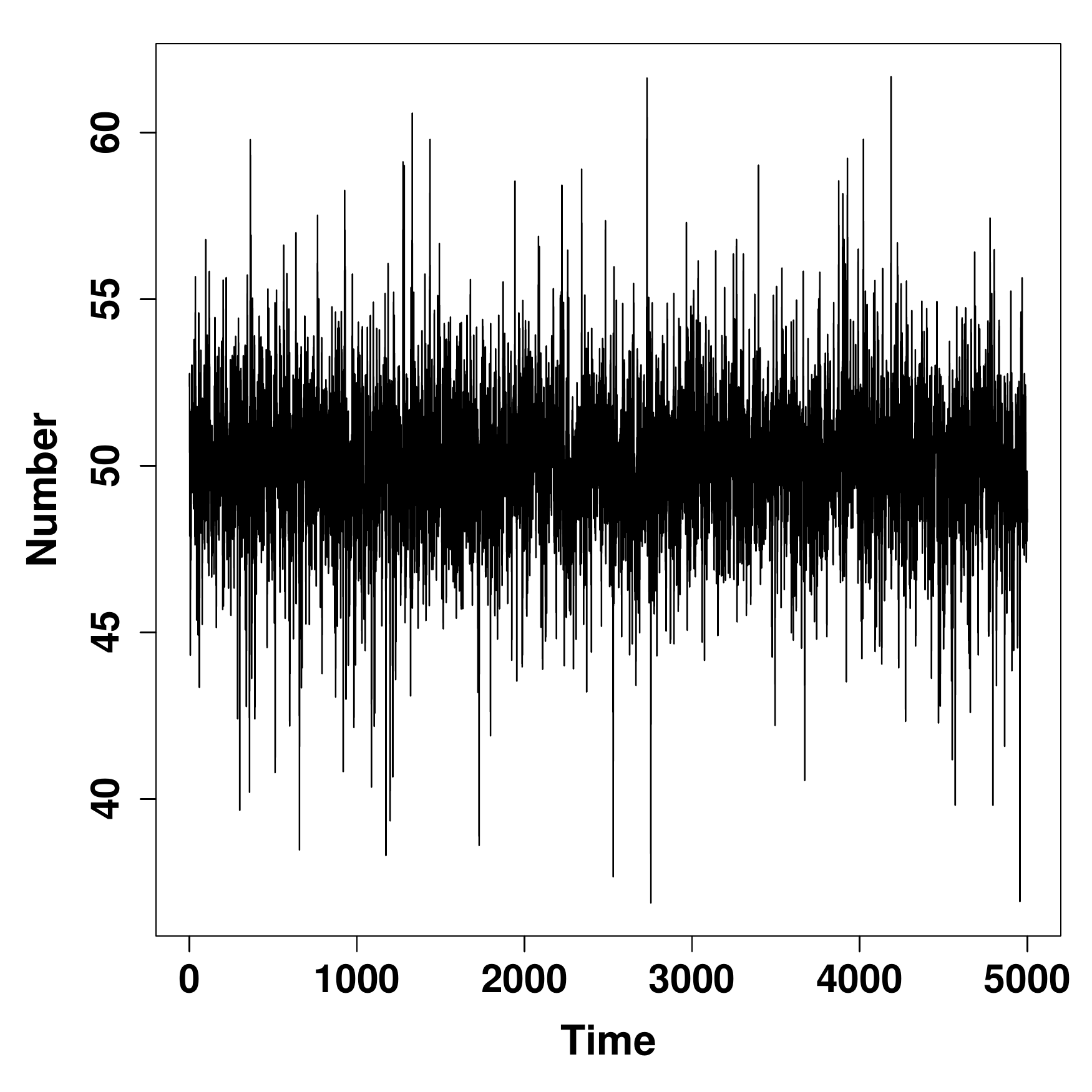}}
\subfigure[Time series 3]{\includegraphics[width=0.32\textwidth]{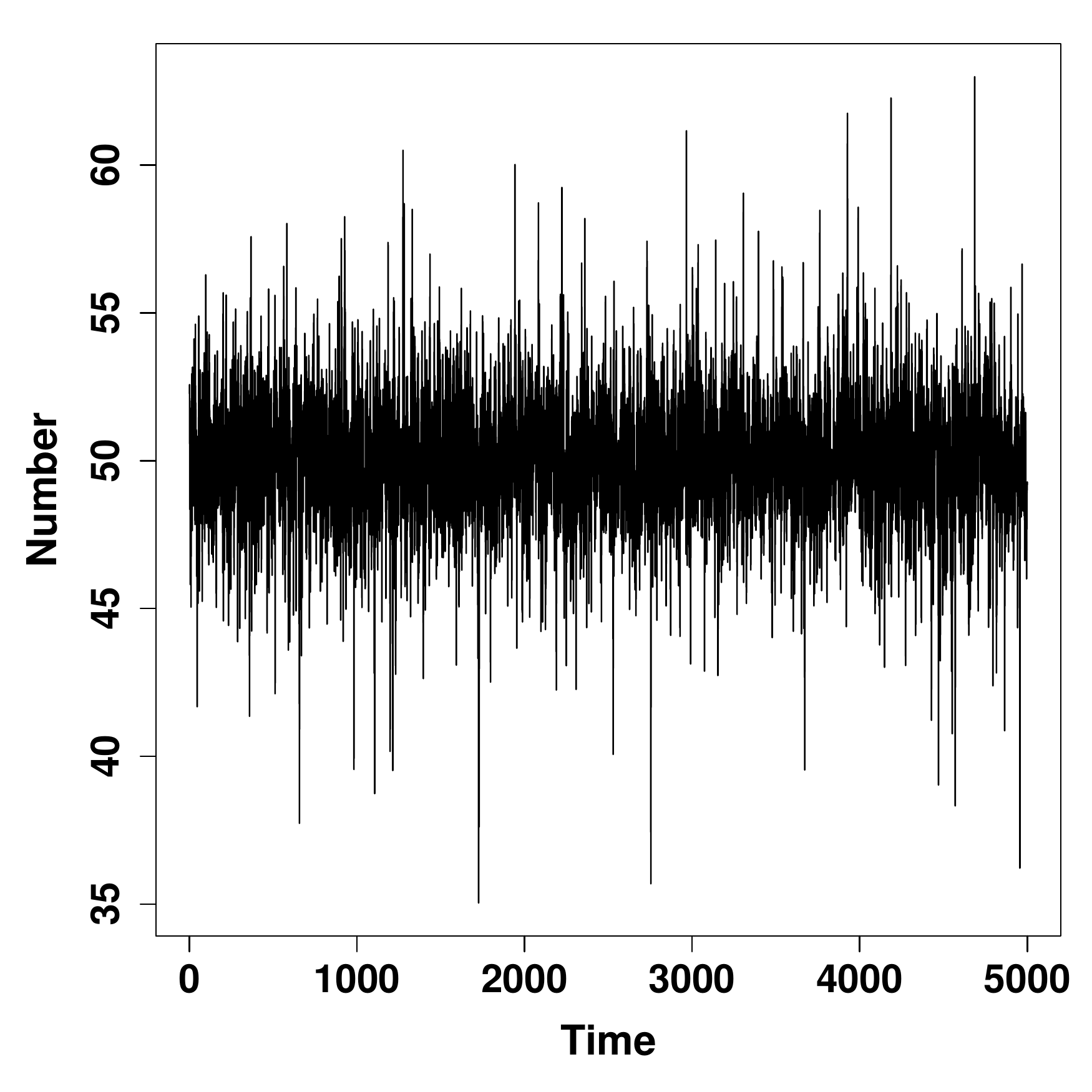}}
\subfigure[Time series 4]{\includegraphics[width=0.32\textwidth]{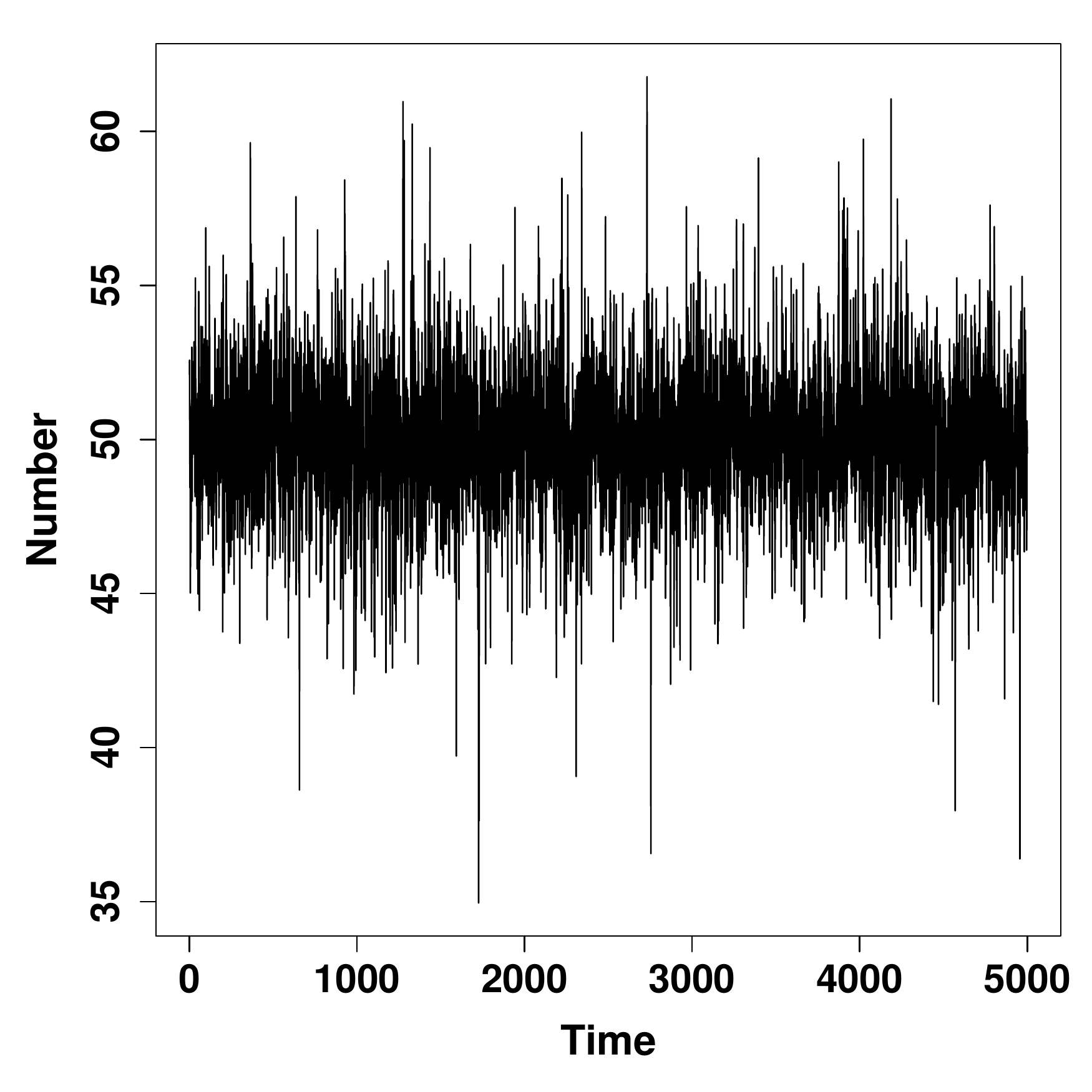}}
\subfigure[Time series 5]{\includegraphics[width=0.32\textwidth]{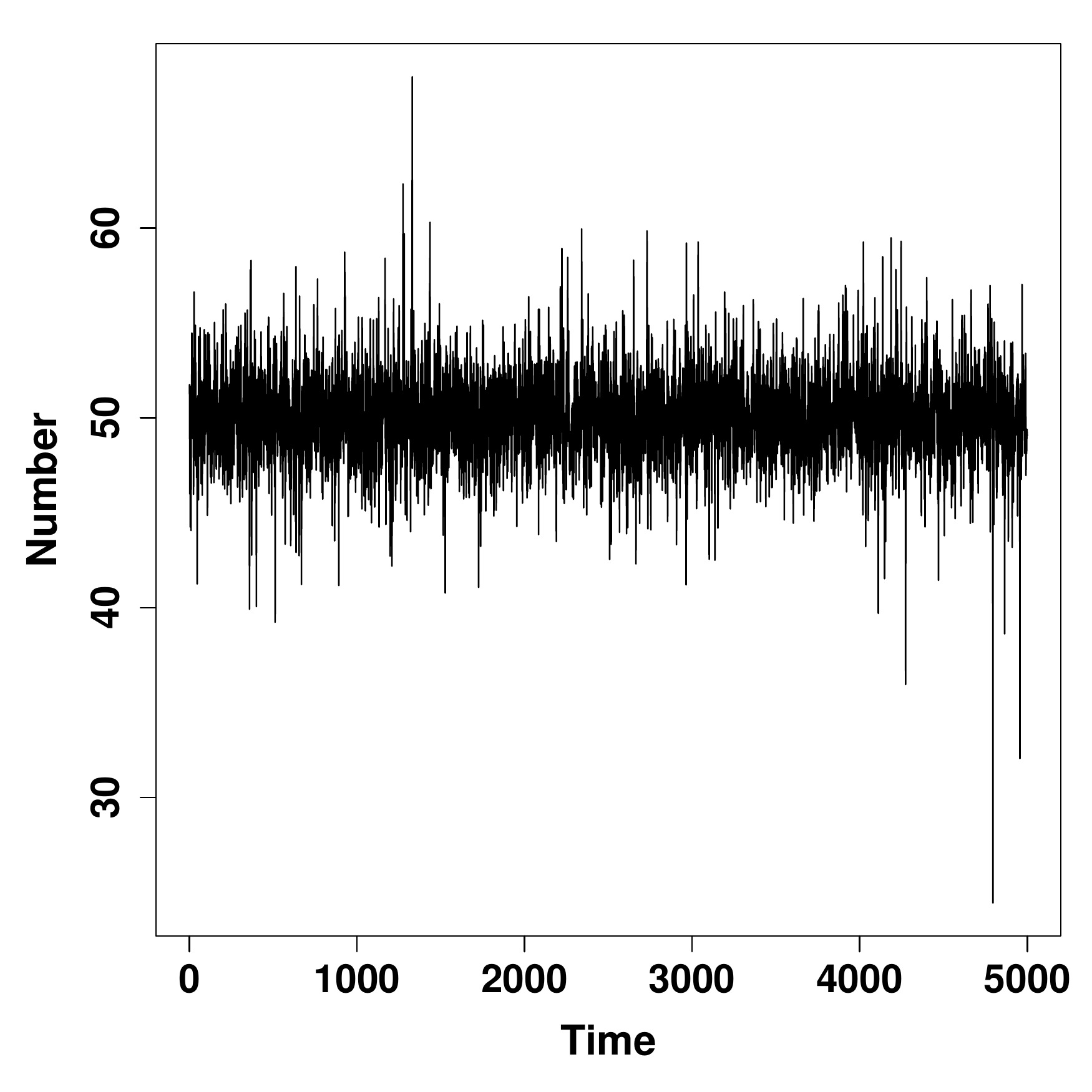}}
\caption{Time series plots of simulated Copula+AR-GARCH with student-t tail. \label{fig:garch-ts}}
\end{figure}
The simulated time series plots are displayed in
Figure \ref{fig:garch-ts}, and the extreme values can also be observed.

We again employ Algorithm \ref{alg1} to train and validate the deep learning model, and the residuals are used to fit the GPD distribution in Eq. \eqref{eq:gpd}. It is discovered that the threshold of $90$th percentile of residuals can produce satisfactory fitting performance for all the time series. Therefore, in the prediction procedure of Algorithm \ref{alg2}, the threshold is fixed at $90$th percentile of residuals for each time series.
\begin{table}[htbp!]\centering
\centering
\caption{Prediction performances based on MAPE and MSE for the proposed model and benmarkmodel. \label{table:var-test2}}
\begin{tabular}{l|cc|cc}
  \toprule
 Series &\multicolumn{2}{|c|}{Deep}   &\multicolumn{2}{|c}{Benchmark}\\ \midrule
 &MAPE&MSE &MAPE&MSE     \\\midrule
    1  & 0.0185 & 2.4612 &  0.032 & 4.8339  \\ \midrule
    2  & 0.0189 & 2.3415 &   0.0324 & 4.6922   \\ \midrule
    3  & 0.0198 & 2.5678 &   0.0325 & 4.7743  \\ \midrule
    4  & 0.0190 & 2.2672 &   0.0313 & 4.3606 \\ \midrule
    5  & 0.0183 & 2.3662 &   0.0323 & 4.7806  \\
 \bottomrule
\end{tabular}
\end{table}
Table \ref{table:var-test2} presents the  performances of point predictions for both proposed model and benchmark model. It is seen that the proposed model can significantly outperform  the benchmark model in terms of both metrics for all the time series.

\begin{figure}[htbp!]
\centering
\subfigure[Violations-Deep+EVT]{\includegraphics[width=0.32\textwidth]{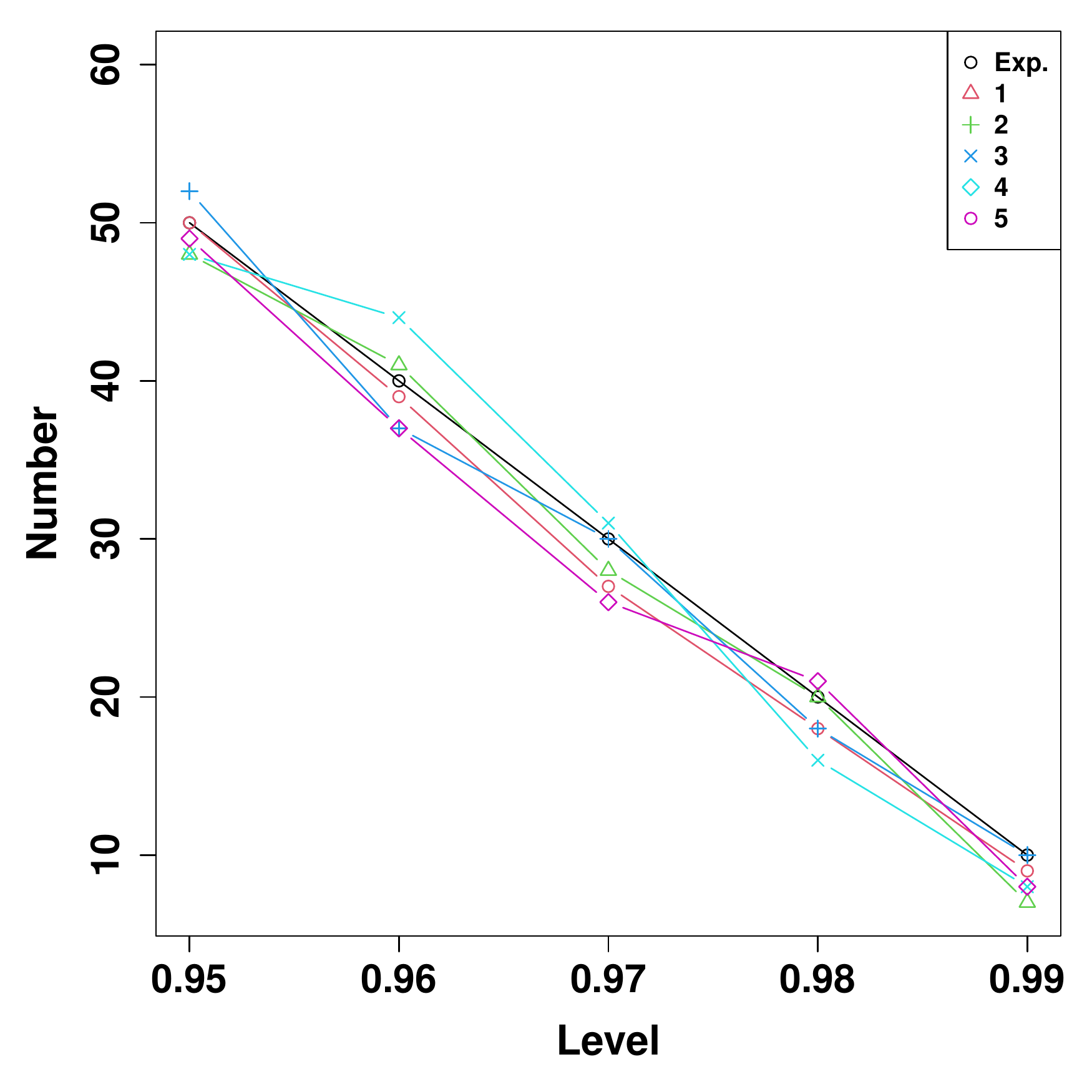}\label{fig:vio-garch-deep}}
\subfigure[LRuc-Deep+EVT]{\includegraphics[width=0.32\textwidth]{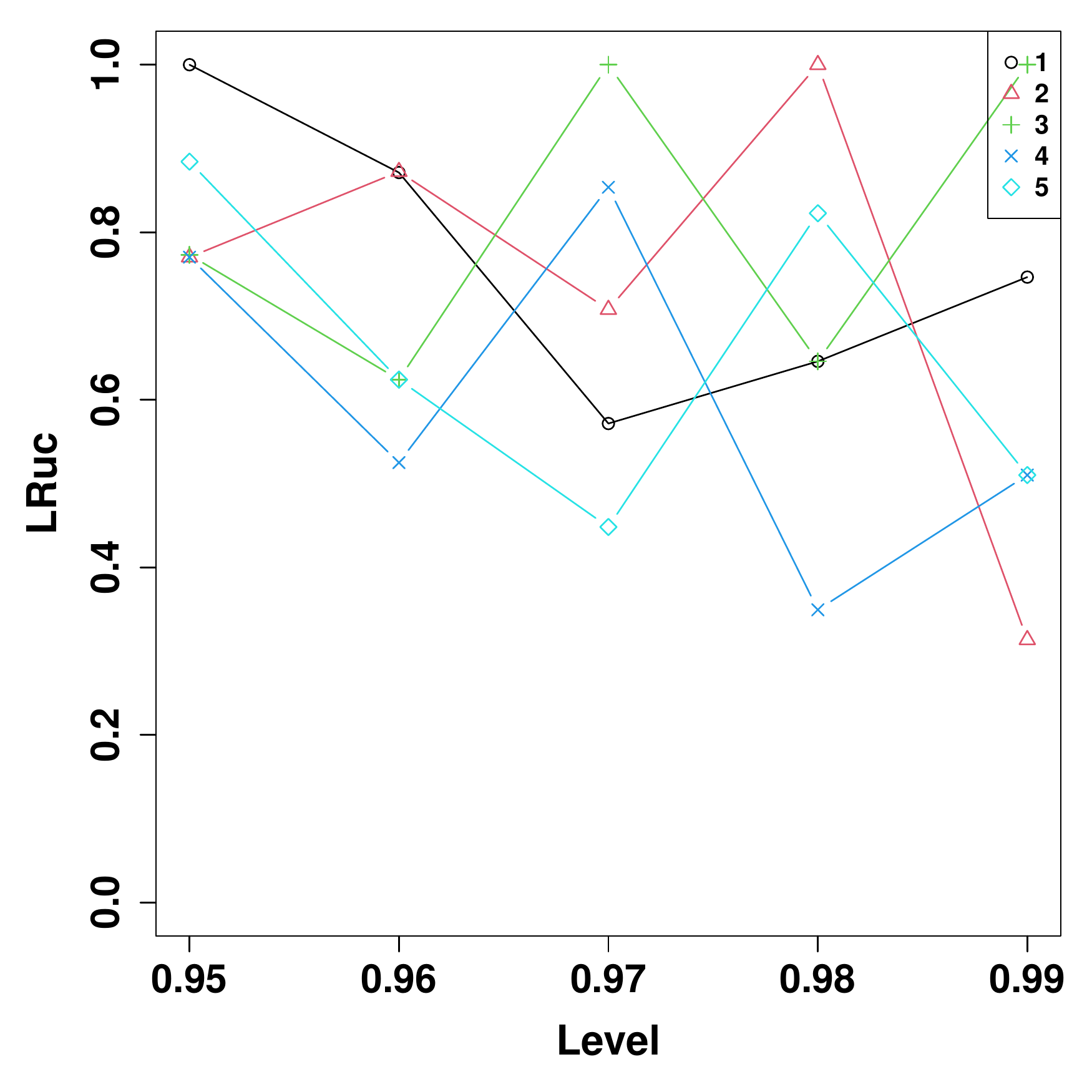}\label{fig:lruc-garch-deep}}
\subfigure[LRcc-Deep+EVT]{\includegraphics[width=0.32\textwidth]{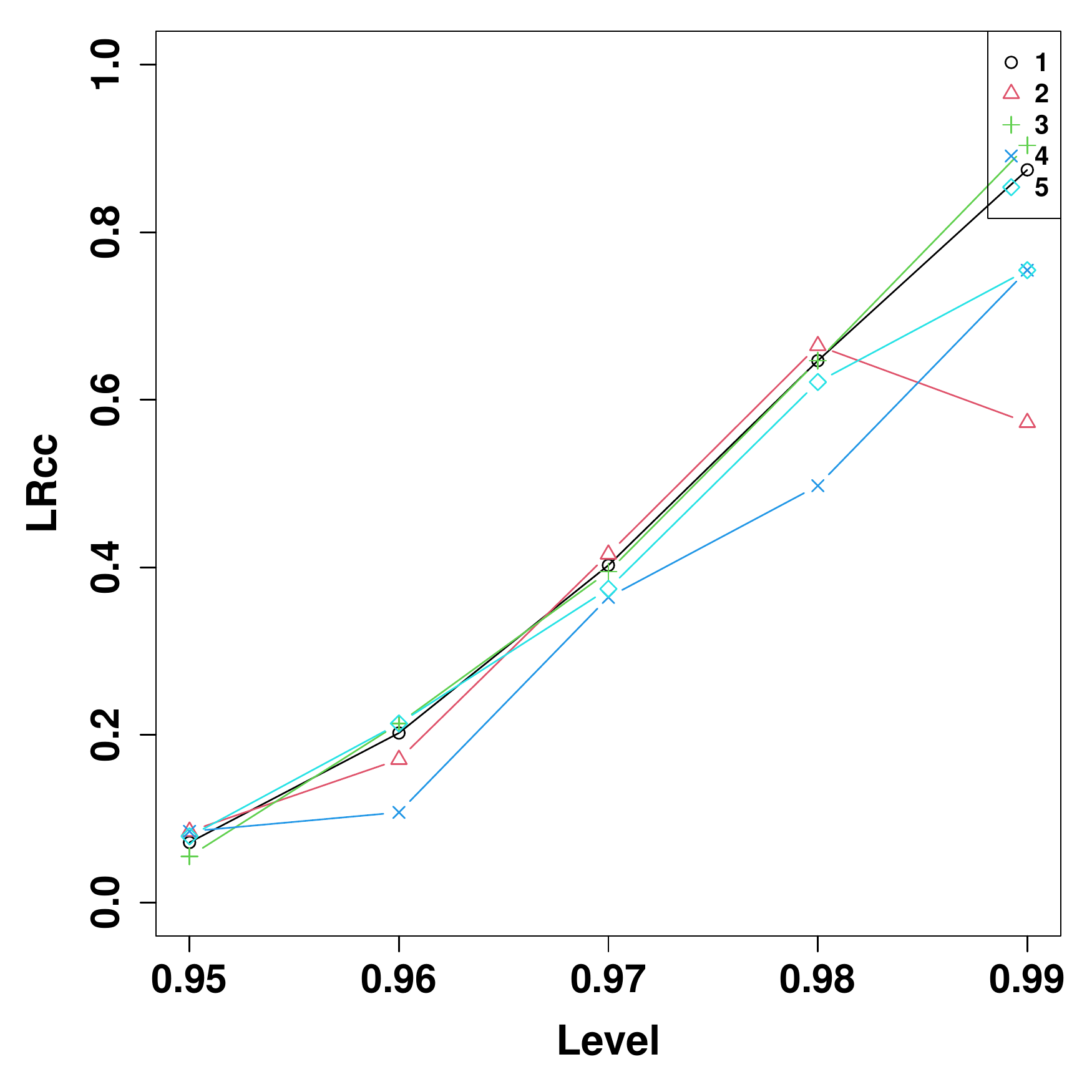}\label{fig:lrcc-garch-deep}}
\subfigure[Violations-Benchmark]{\includegraphics[width=0.32\textwidth]{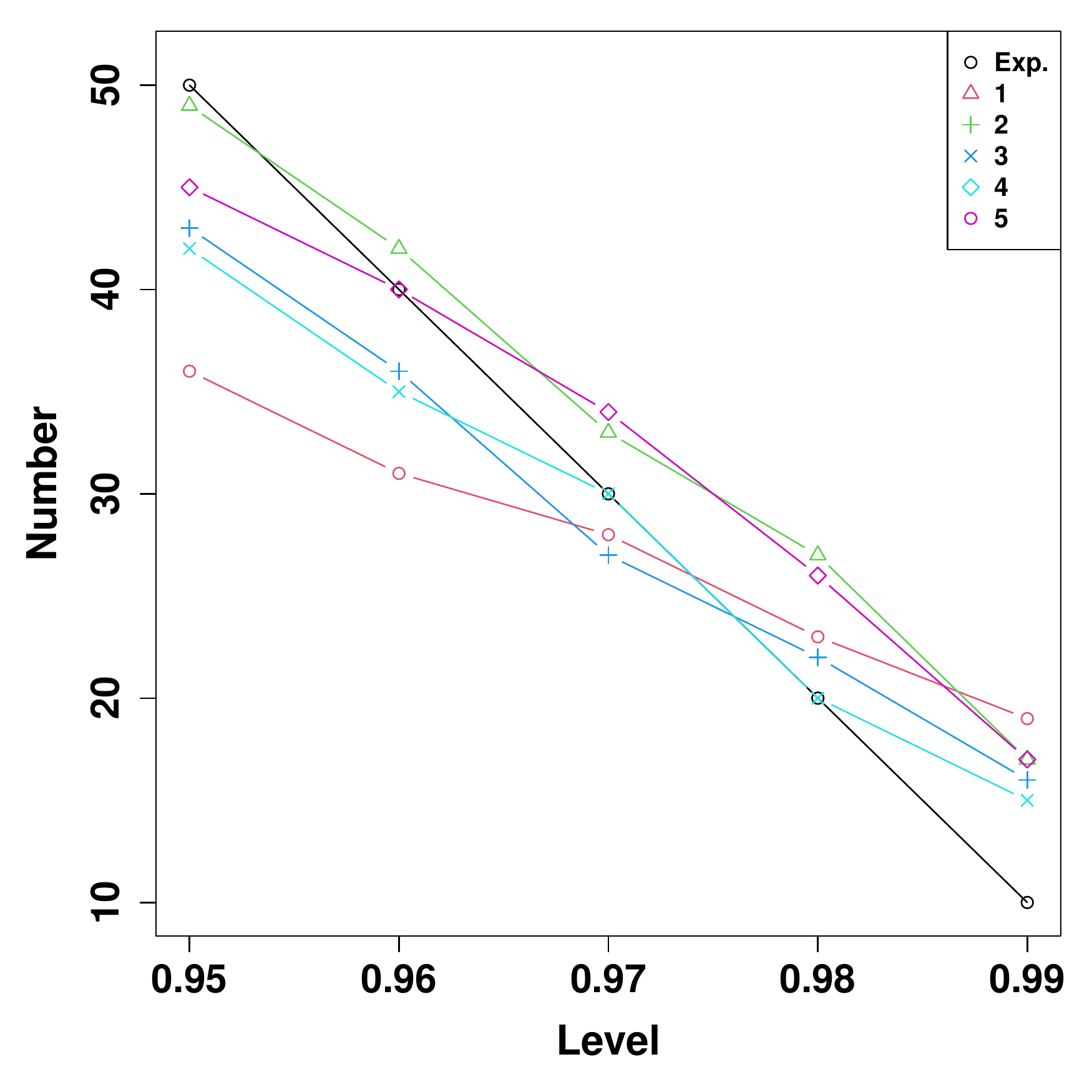}\label{fig:vio-garch-ben}}
\subfigure[LRuc-Benchmark]{\includegraphics[width=0.32\textwidth]{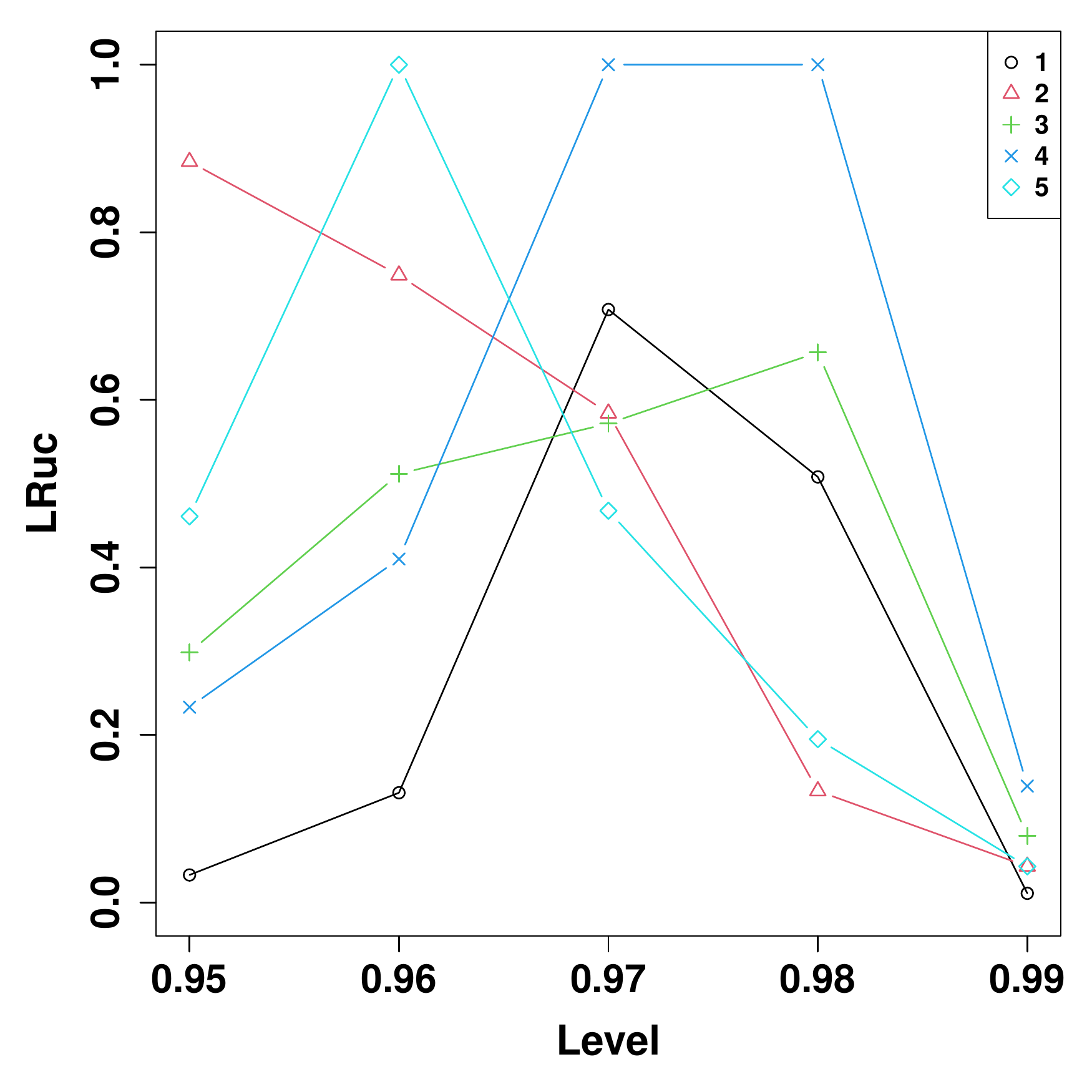}\label{fig:lruc-garch-ben}}
\subfigure[LRc-Benchmark]{\includegraphics[width=0.32\textwidth]{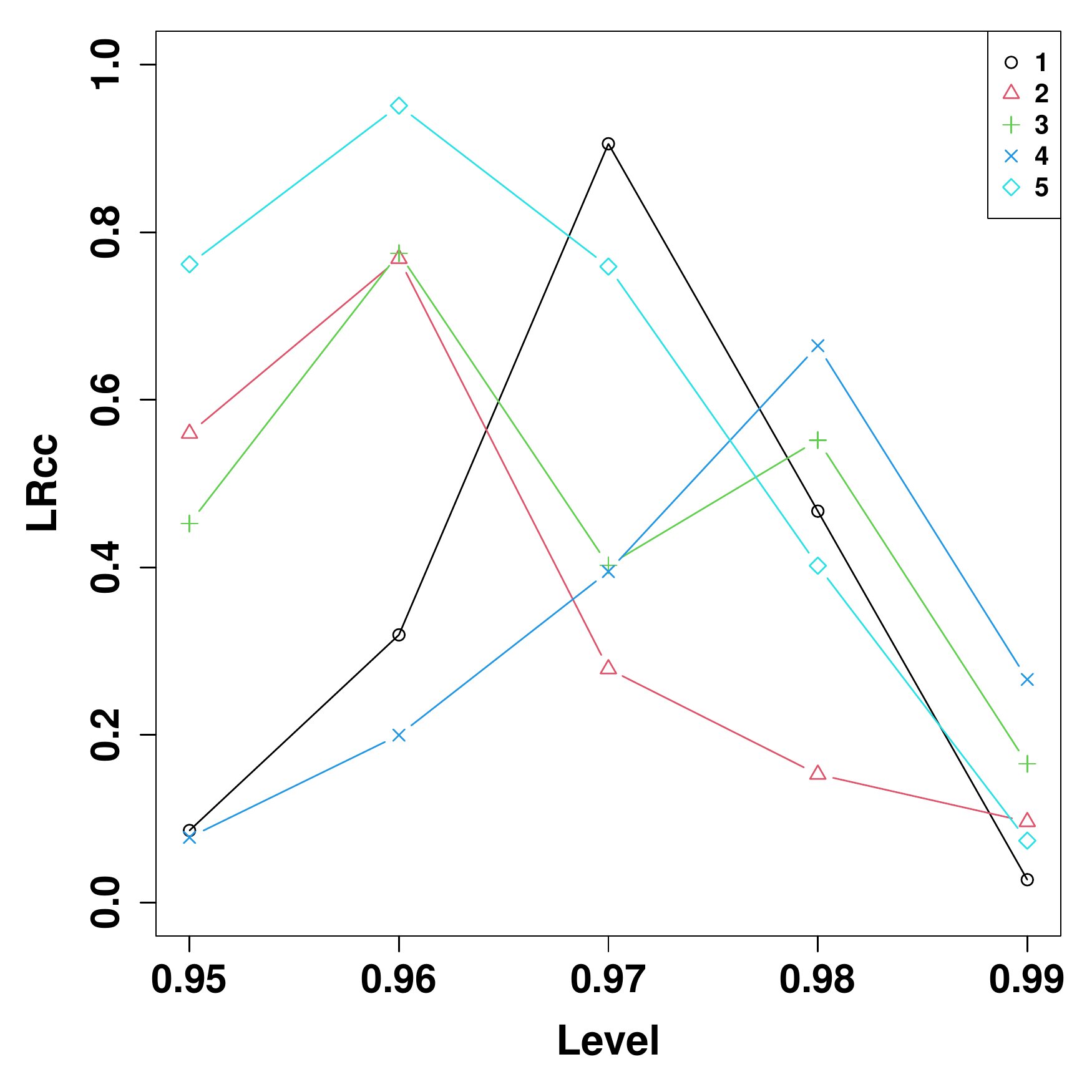}\label{fig:lrcc-garch-ben}}
\caption{Numbers of violations and $p$-values of LRuc and LRcc tests for proposed model,  and true model. \label{fig:garch-data2-violation}}
\end{figure}
For the high quantiles, the number of violations are displayed in Figures \ref{fig:vio-garch-deep} and \ref{fig:vio-garch-ben}. It is seen that the proposed model has a very satisfactory prediction performance compared to that of benchmark model. The large $p$-values of LRuc and LRcc tests in Figures \ref{fig:lruc-garch-deep} and \ref{fig:lrcc-garch-deep} compared to those in \ref{fig:lruc-garch-ben} and \ref{fig:lrcc-garch-ben} also confirm that the proposed approach has an accurate prediction  performance for the high quantiles.
  \end{itemize}
To conclude, the proposed model has satisfactory prediction performances for both the point prediction and high quantile prediction. It can significantly outperform the benchmark model.

 \section{Empirical study}\label{sec:empirical}
  In this section, we study two real attack data which are collected by the honeypot instrument \cite{spitzner2003honeynet}.
 \subsection{Honeypot data I with 9 dimensions}
 This honeypot data is publicly available on the web \cite{JR2019}, which was collected via Amazon Web Service (AWS) virtual honeypots across the world. The dataset has 9 honeypot hosts (EU, Oregon, Singapore, SA, Tokyo, Norcal1, Norcal2, US-East, Sydney), and the attacks were recorded between 03/03/2013 to 09/08/2013, which includes 451,581 events. The recorded attack data includes attack
time, targeted host, attackers’ IP addresses and origin countries. This dataset  was discussed in \cite{ling2019predicting} based on the daily aggregation, where the long range dependence model was discovered and incorporated into the modeling process.

\paragraph{Daily aggregation.} The daily aggregated data has 188 observations,  we leave the last 38 observations as the prediction evaluation as that in \cite{ling2019predicting} where the predicted MSE is reported. We employ Algorithm
\ref{alg1} to build the deep learning model where the training data has 100 observations, and validation data has 50 observations. The prediction performance based on Algorithm \ref{alg2} is reported in Table \ref{table:day-honeypot}.
\begin{table}[htbp!]\centering
\centering
\caption{Prediction performances for log-transformed daily aggregated honeypot attack data.  \label{table:day-honeypot}}
\begin{tabular}{l|cc|cc}
  \toprule
 Series &\multicolumn{2}{|c|}{Deep}   &\multicolumn{1}{|c}{LRD+VAR in \cite{ling2019predicting} }\\ \midrule
 &MAPE&MSE  &MSE   \\\midrule
EU   &    0.0685 &0.2058   & 0.2540  \\\midrule
 Oregon    &   0.0118 & 0.0096   & 0.0138   \\\midrule
 Singapore    &  0.0144 &0.0145    & 0.0133 \\  \midrule
 SA    &  0.0284 & 0.0319  & 0.0567 \\  \midrule
  Tokyo   &0.0418 &0.4225    & 0.4733 \\   \midrule
Norcal1    & 0.0262 &0.0282   &  0.0445  \\   \midrule
Norcal2   & 0.0287 & 0.0358  &0.0692 \\   \midrule
US-East  &  0.0666 &0.3726  &  0.3848  \\   \midrule
Sydney   &   0.0554 & 0.1126  & 0.1374  \\
 \bottomrule
\end{tabular}
\end{table}
The proposed deep learning has a satisfactory prediction performance. In particular, it outperforms the approach in  \cite{ling2019predicting} for all the attack time series except for Singapore based on MSEs (0.0145 vs 0.0133). Since the daily aggregated time series only has a total of 188 observations for each host, the high quantile predictions are not performed.

\begin{figure}[htbp!]
\centering
\subfigure[EU]{\includegraphics[width=0.32\textwidth]{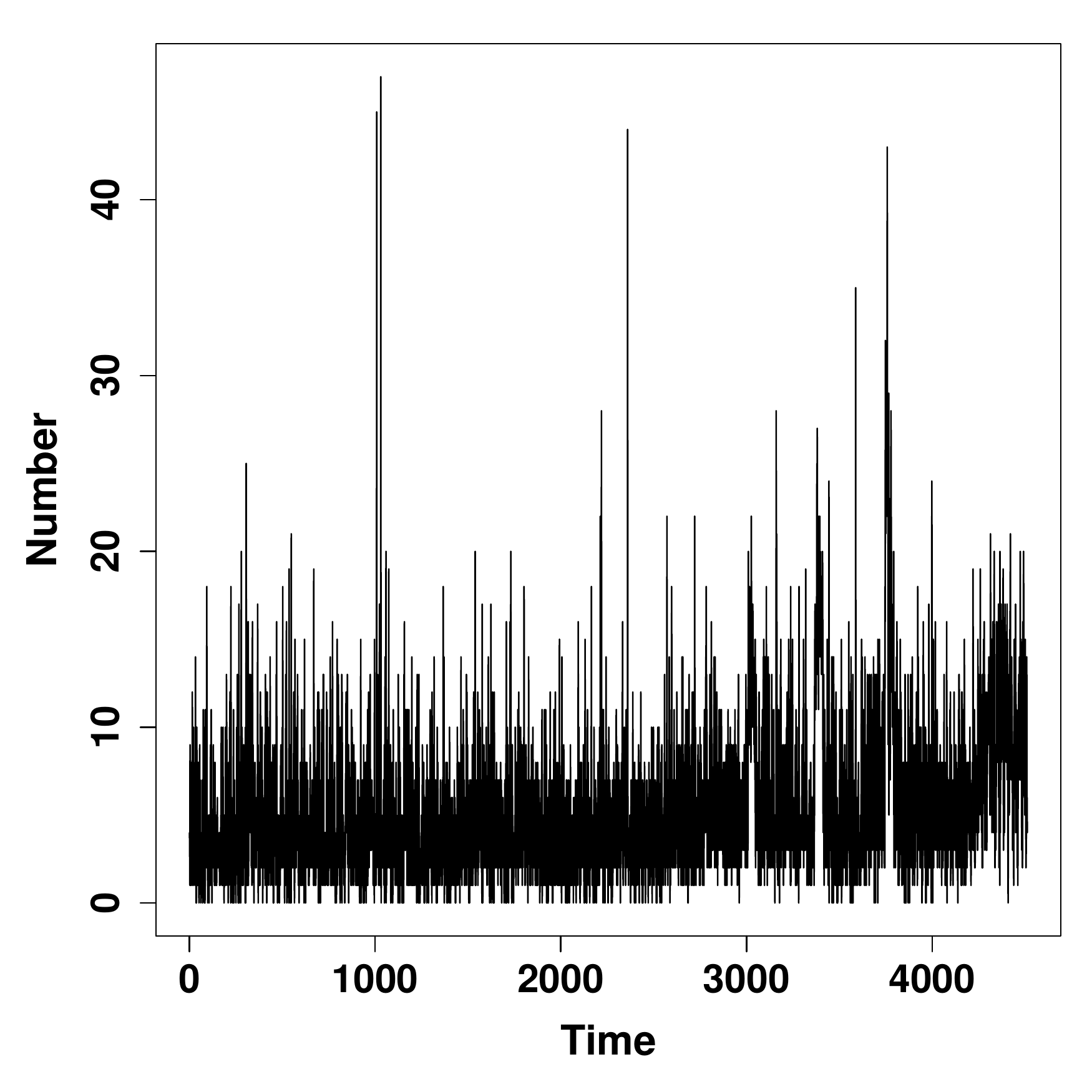}}
\subfigure[Oregon]{\includegraphics[width=0.32\textwidth]{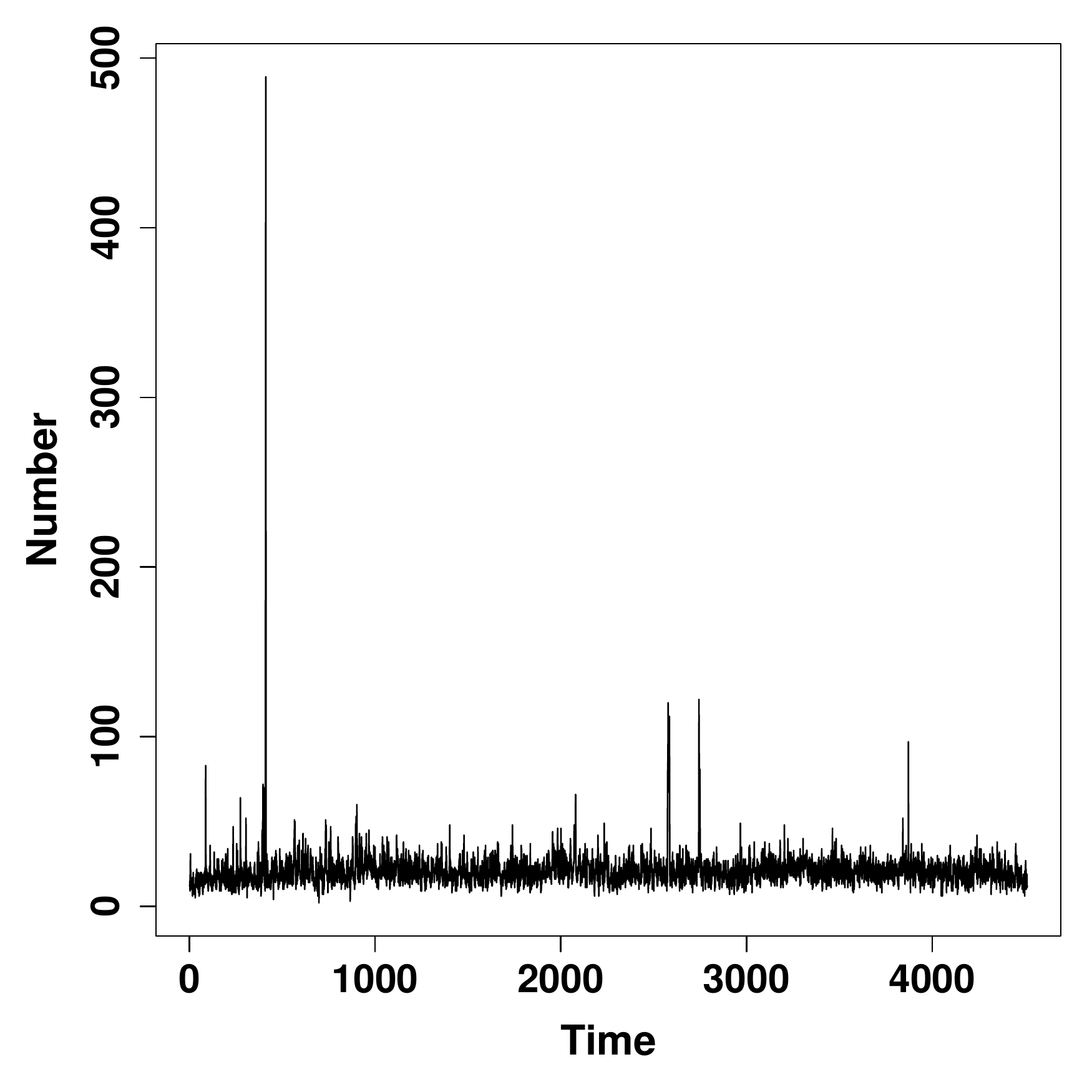}}
\subfigure[Singapore]{\includegraphics[width=0.32\textwidth]{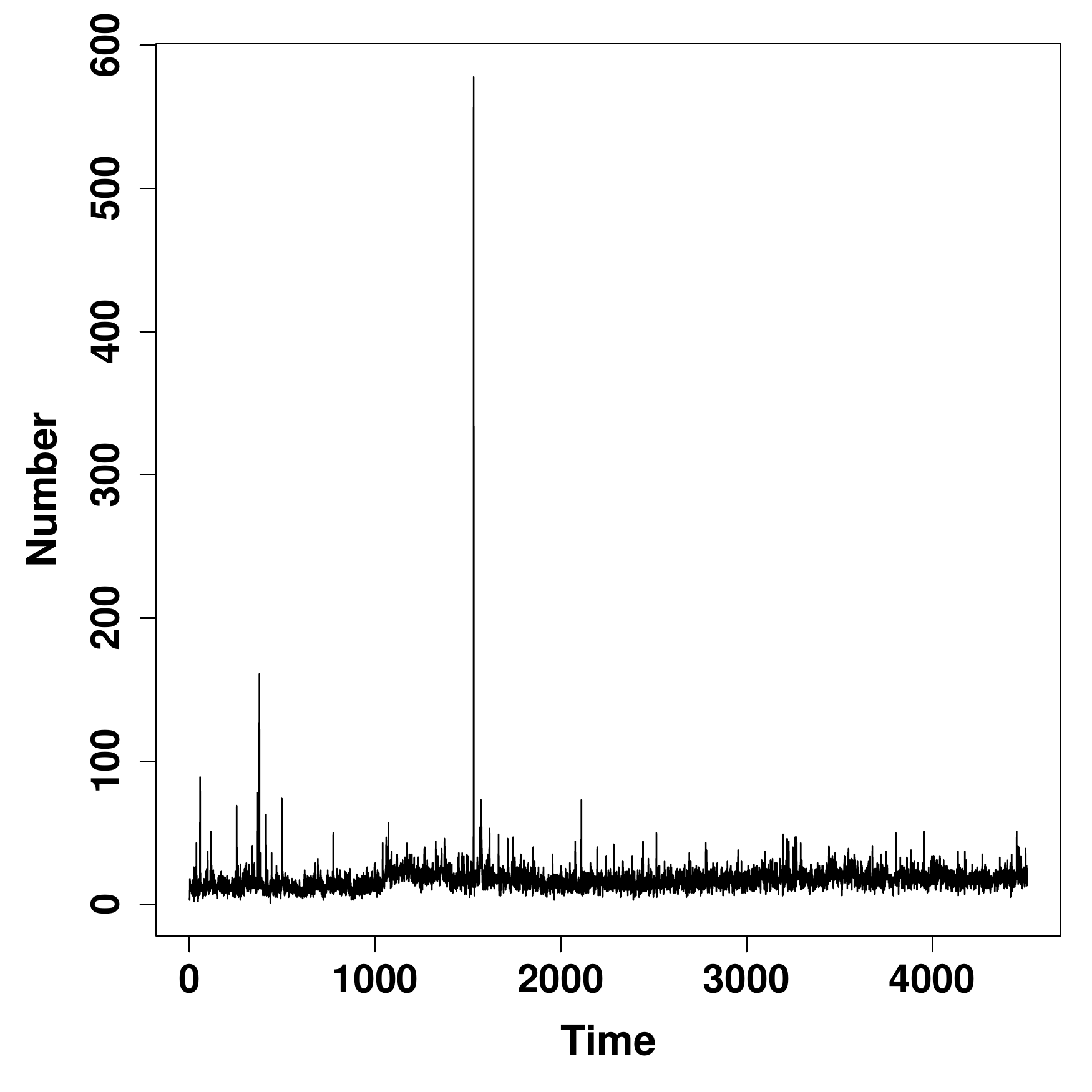}}
\subfigure[SA]{\includegraphics[width=0.32\textwidth]{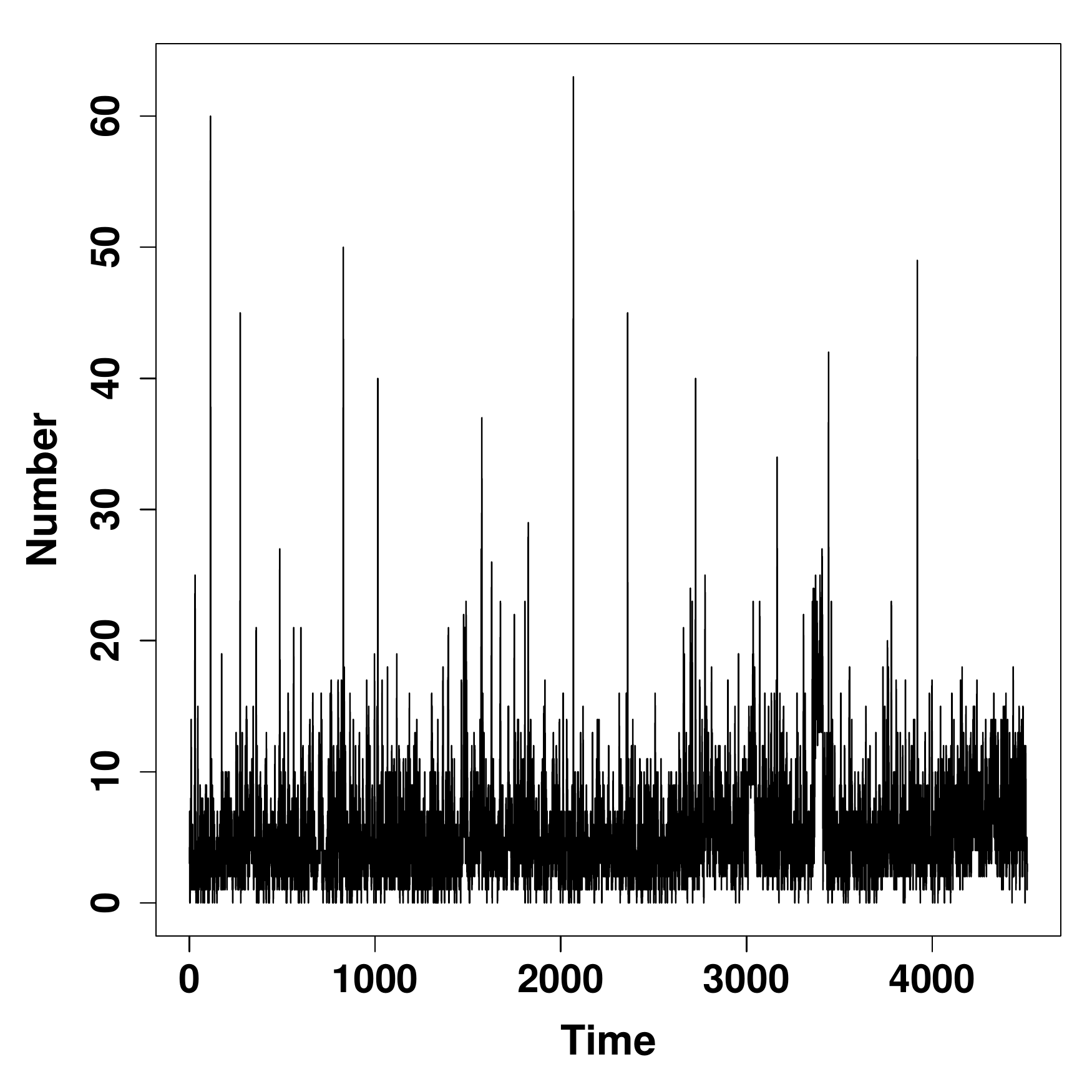}}
\subfigure[Tokyo]{\includegraphics[width=0.32\textwidth]{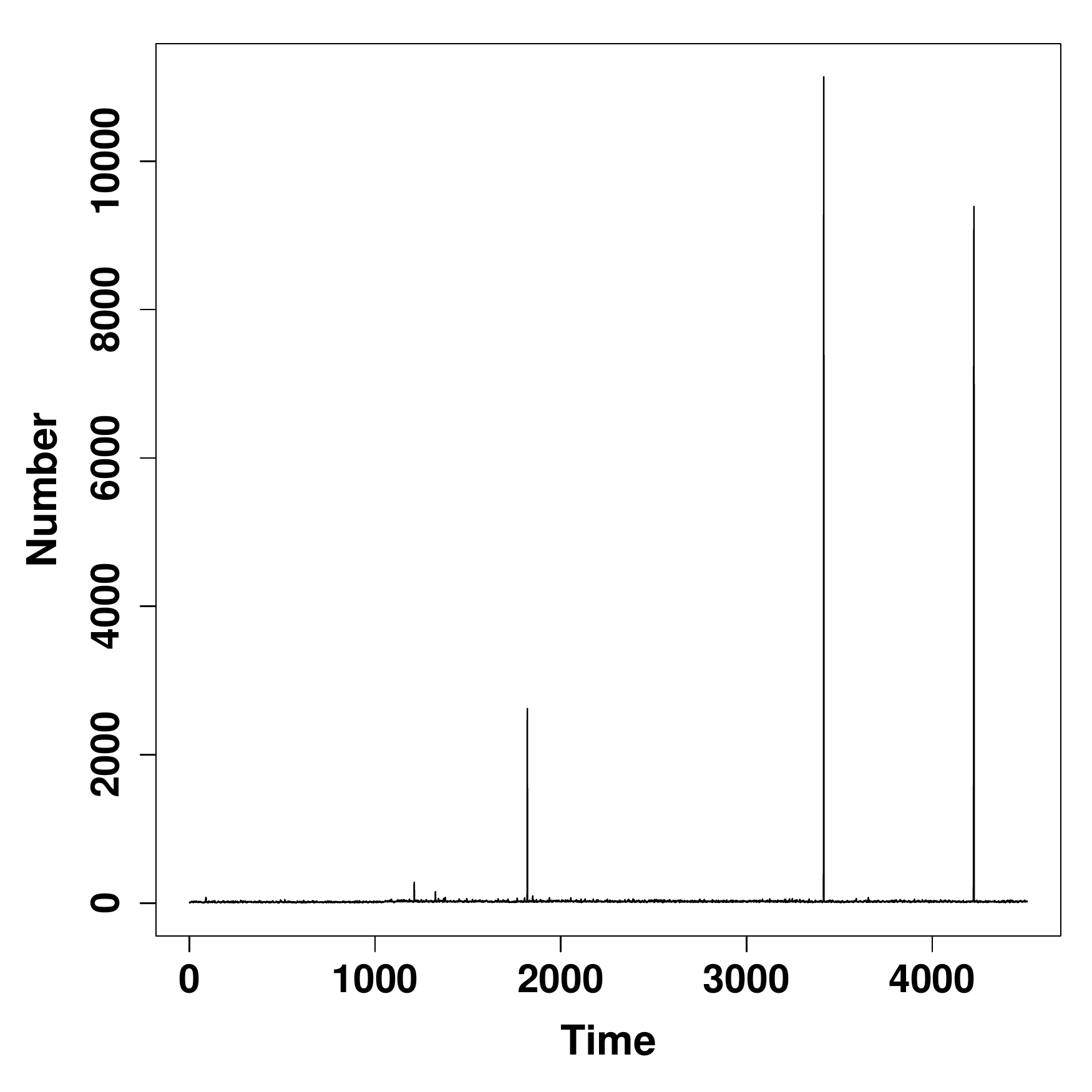}}
\subfigure[Norcal1]{\includegraphics[width=0.32\textwidth]{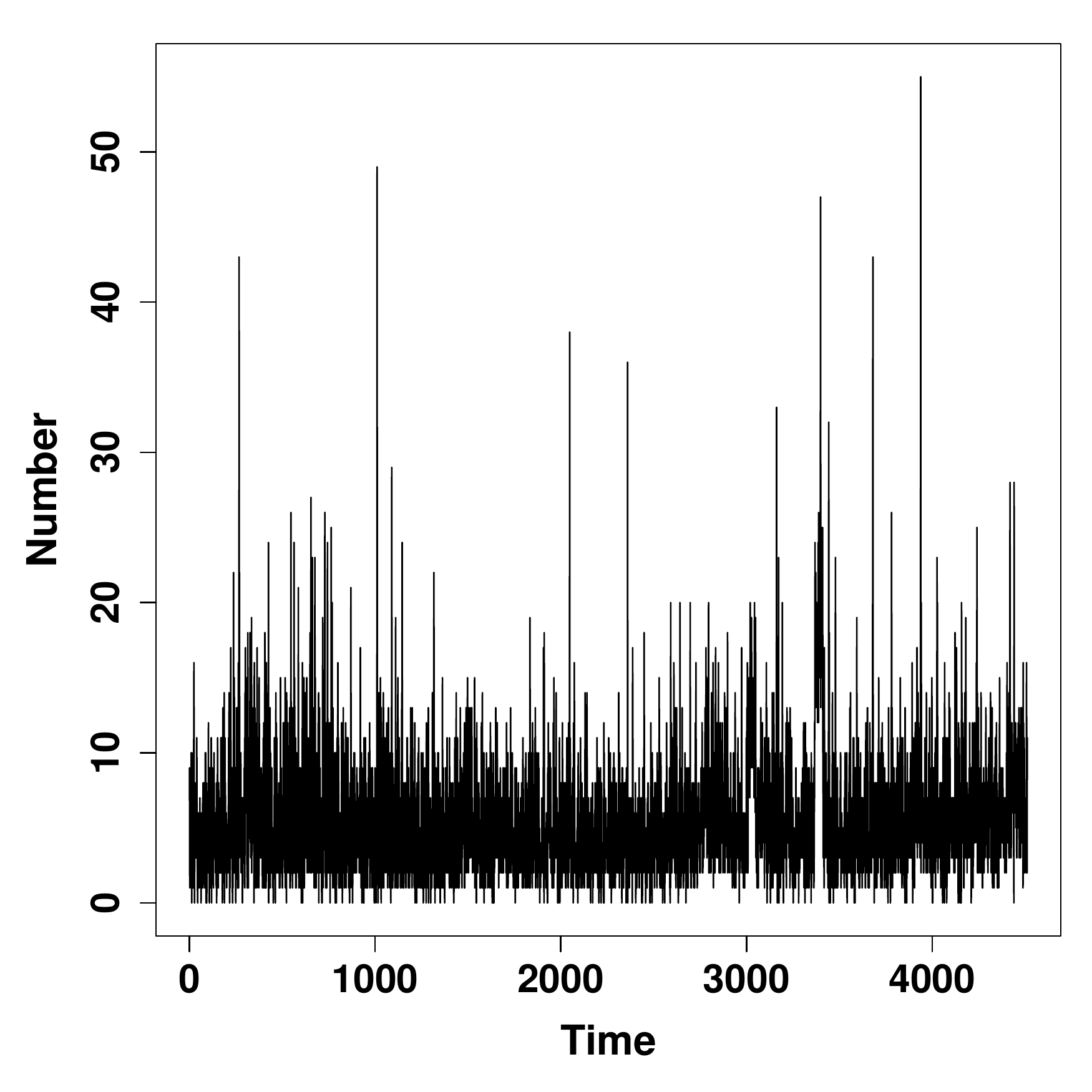}}
\subfigure[Norcal2]{\includegraphics[width=0.32\textwidth]{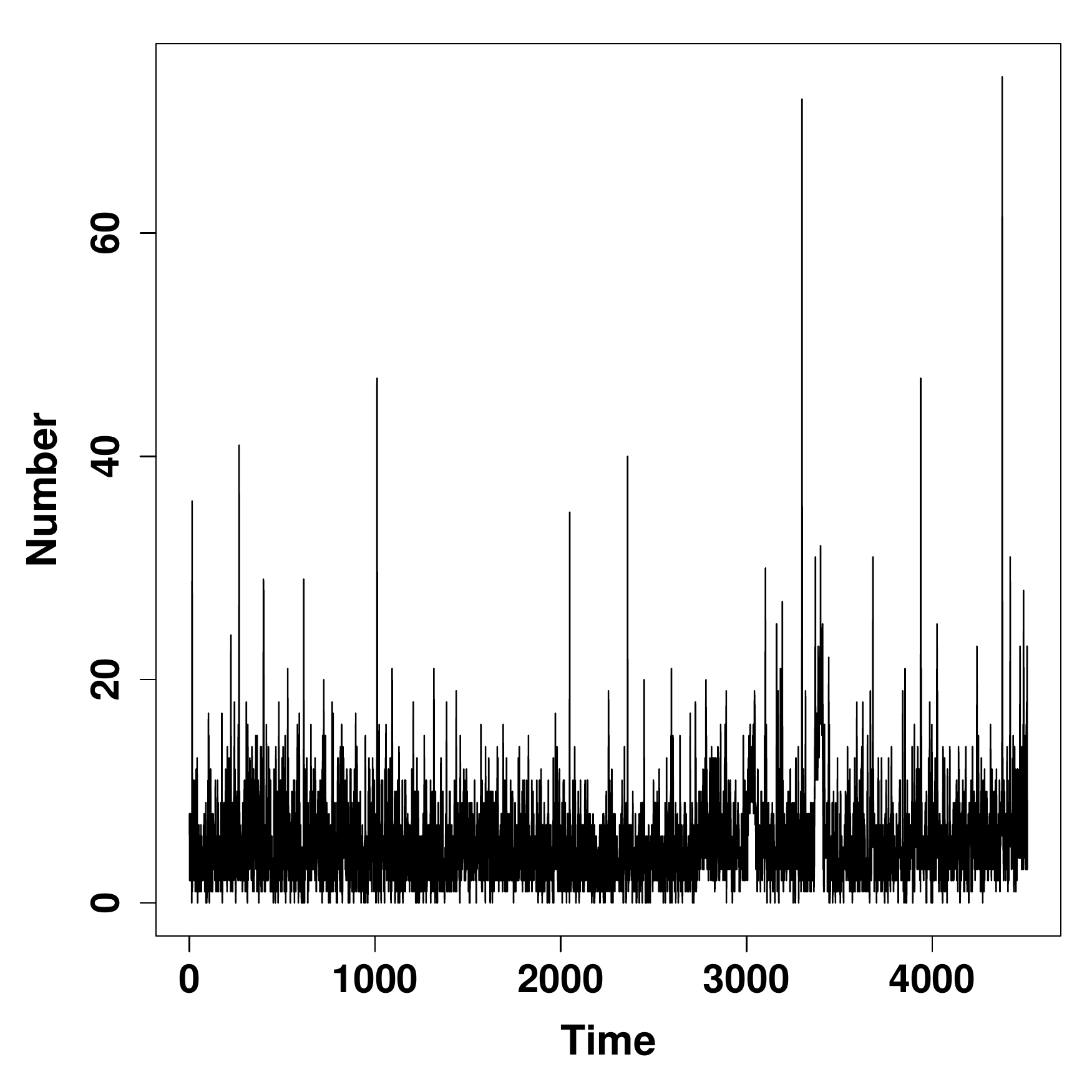}}
\subfigure[US-East]{\includegraphics[width=0.32\textwidth]{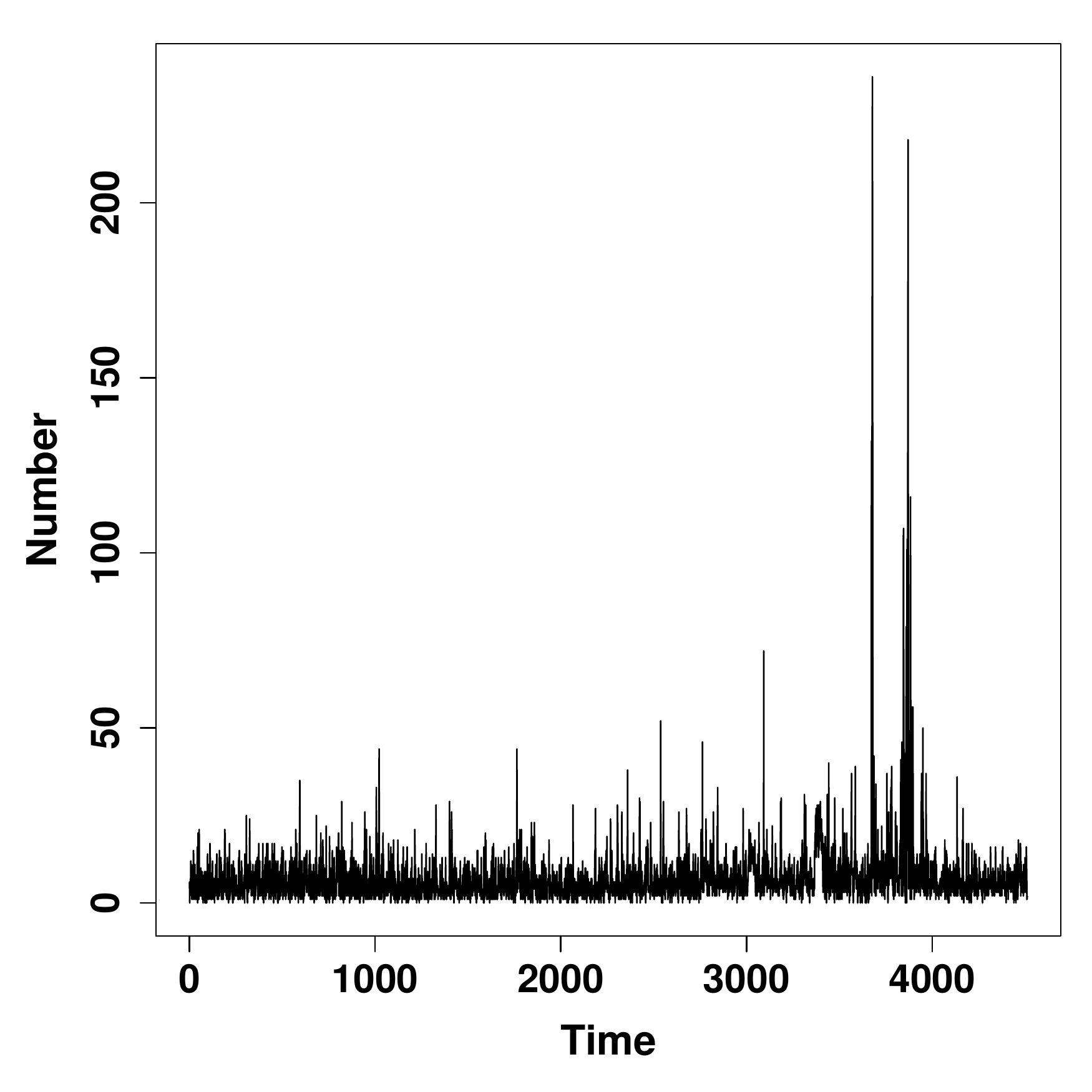}}
\subfigure[Sydney]{\includegraphics[width=0.32\textwidth]{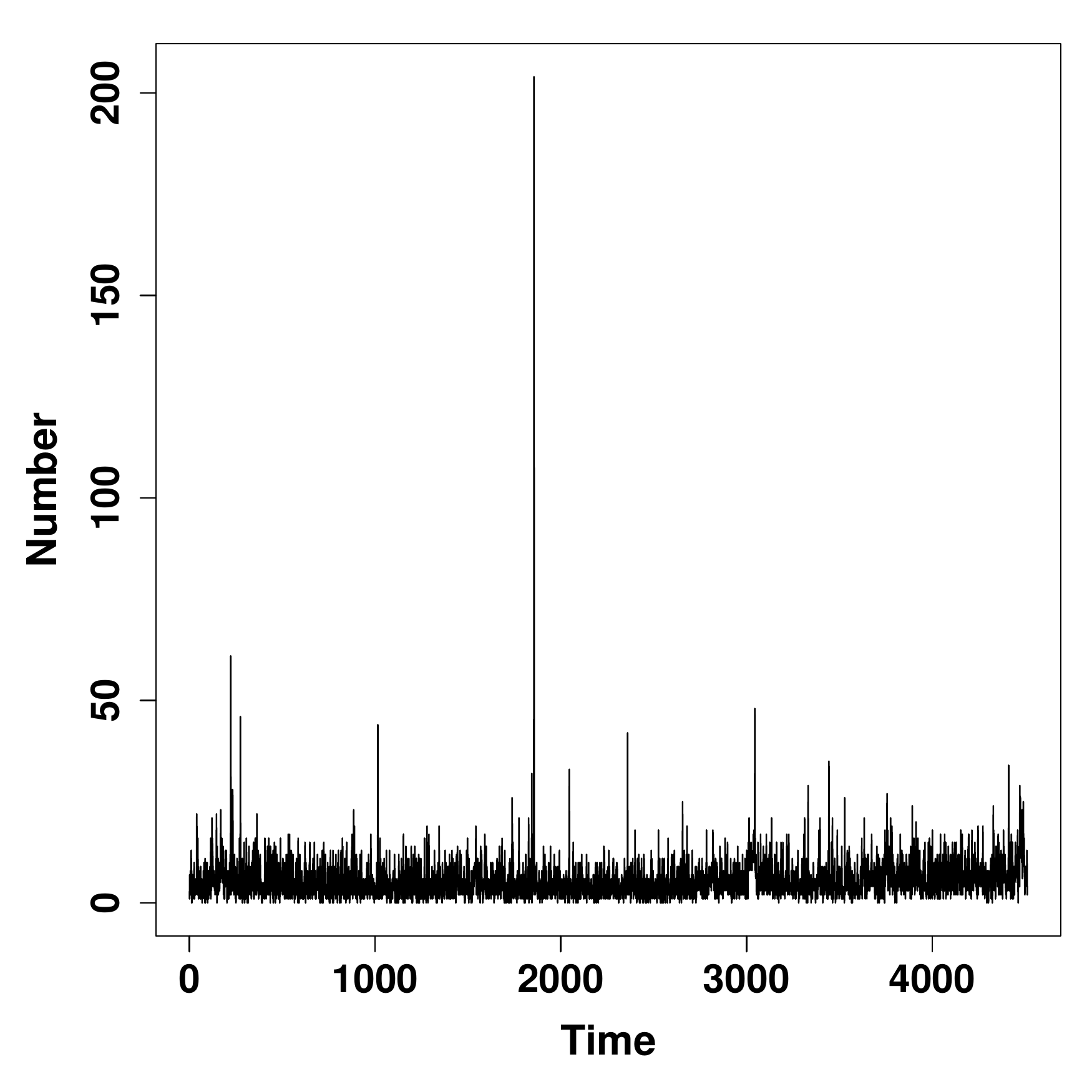}}
\caption{Time series plots of hourly aggregated  attack data for all the hosts. \label{fig:attack-ts}}
\end{figure}

\paragraph{Hourly aggregation.} In practice, the network defender is often interested in assessing the hourly attacks \cite{zhan2013characterizing,zhan2015predicting,peng2018modeling}. In this section, we study the performance of proposed model based on the hourly aggregated data which has 4512 observations for each host. The hourly aggregated time series are plotted in Figure \ref{fig:attack-ts}. It is observed that there are extreme attacks for all the hosts. Particularly, there exist tremendous numbers of attacks during some short periods  for Oregon, Singapore, and Tokyo hosts. The data is split into three parts: training with { 3500} observations, validation with {500} observations, and the last 512 observations are used for the prediction evaluation.
The Algorithm \ref{alg1} is employed on the training and validation datasets to select the best deep learning model, and Algorithm \ref{alg2} is used for the prediction.

The prediction performance is reported in Table \ref{table:honeypot1-test}. For the comparison purpose, the prediction performance of benchmark VAR model is also reported. For the proposed approach, we can observe that it can significantly outperform the benchmark model in terms of MAPE metric. For the MSE metric, it is seen that the proposed approach can also outperform the bencmark model for all the time series except for SA (23.7633 vs 14.6417). For Tokyo time series, it is seen that the MSEs are very large for both models. This is mainly because of one extreme observation which can not be predicted well.
\begin{table}[htbp!]\centering
\centering
\caption{Prediction performances for daily aggregated honeypot attack data.   \label{table:honeypot1-test}}
\begin{tabular}{l|cc|cc}
  \toprule
 Series &\multicolumn{2}{|c|}{Deep}   &\multicolumn{2}{|c}{Benchmark}\\ \midrule
 &MAPE&MSE &MAPE&MSE   \\\midrule
EU& 0.3067   &   15.5700  & 0.4055 & 15.7607  \\\midrule
 Oregon &0.1826  &    34.6819 & 0.2446 & 35.9174  \\\midrule
Singapore & 0.1756  &    29.5333 & 0.2372 & 34.3774  \\  \midrule
SA& 0.3248   &   23.7633 & 0.4643 & 14.6417 \\  \midrule
  Tokyo&0.6727  &230565.6248 &1.0707 & 321396.0317 \\   \midrule
Norcal1&0.3225   &   12.0416 & 0.4668 & 16.1691   \\   \midrule
Norcal2& 0.3568  &    23.3879 & 0.4857 & 25.7881 \\   \midrule
US-East& 0.4124  &    12.4906 & 0.5408 & 18.9909  \\   \midrule
Sydney& 0.2916  &    13.6534 & 0.4649 & 21.834 \\
 \bottomrule
\end{tabular}
\end{table}

\begin{figure}[htbp!]
\centering
\subfigure[QQ-plot]{\includegraphics[width=0.45\textwidth]{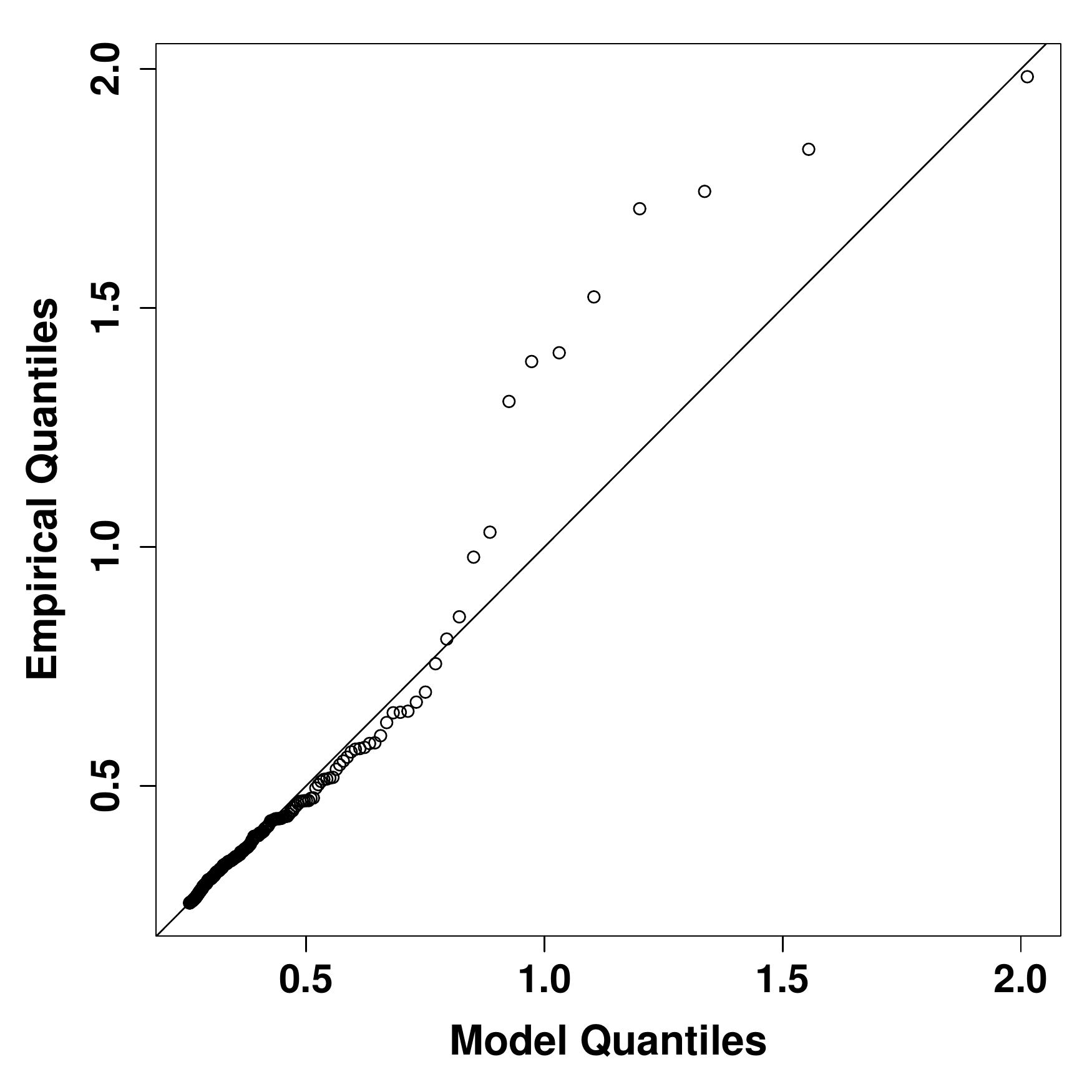}}
\subfigure[PP-plot]{\includegraphics[width=0.45\textwidth]{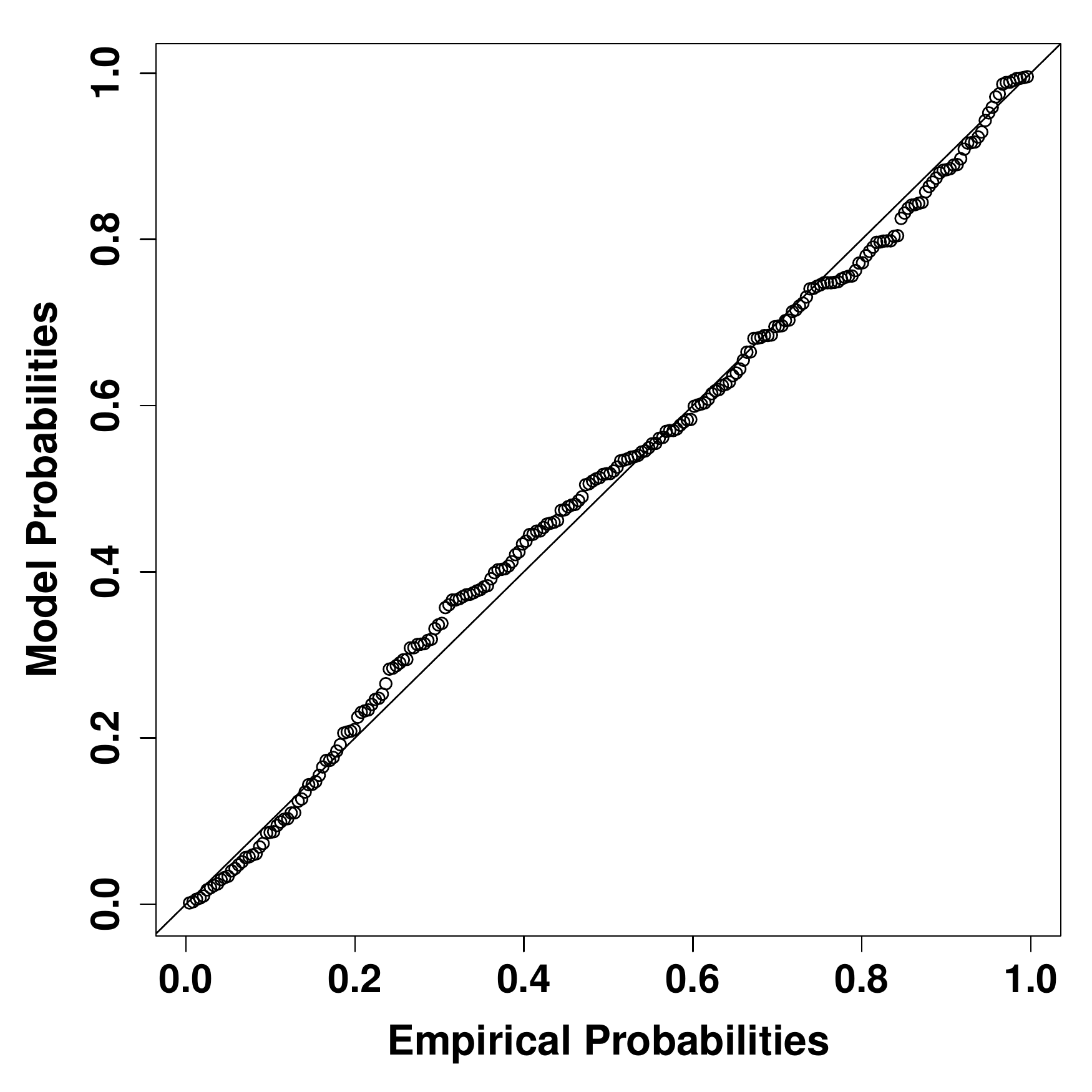}}
\caption{QQ-  and pp- plots of honeypot data fitted by the GPD distribution. \label{fig:honepot-data1-fngpd}}
\end{figure}

For the high quantile prediction, we study the proposed model at the aggregate level. Specifically, we take the log-transformation on both the observations and aggregated fitted values on the training and validation dataset, and calculate the residuals,
$$e_t=\log\left(s_t\right)-\log\left(\hat s_t \right),$$
where $s_t=\sum_{i=1}^{t} y_i$,   $\hat s_t=\sum_{i=1}^{t} y_i$, and $t=1,\ldots,4000$. The GPD distribution is then fitted to $\{e_t\}_{t=1,\ldots,4000}$ with the threshold of $.9$-quantile of $\{e_t\}_{t=1,\ldots,4000}$. The estimated parameters are $\xi=0.3710$ with standard error $0.0827$, and $\sigma=0.0981$ with standard error $0.0100$, which are significant. The QQ- and PP- plots are displayed in Figure \ref{fig:honepot-data1-fngpd}. It is seen that PP- plot is satisfactory. However, the tails of QQ-plot is off the diagonal line, which is mainly caused by the extreme large values in the attack data. Although there are several points off the diagonal line, we still use the fitted GPD distribution to predict the high quantiles as it does not affect the main conclusion.

Algorithm \ref{alg2} is employed on the attack data at the aggregated level. The assessment of predicted $\hat s_t$, $t=4001,\ldots, 4512$ is shown in Table \ref{table:var-test-honeypot1}. It is seen that the prediction performance is very satisfactory. The predicted numbers of violations are quite close to the true numbers of violations. They pass all the tests based on LRuc and LRcc tests except that at $.98$ level of LRcc test is relatively small (0.0241).

\begin{table}[htbp!]\centering
\centering
\caption{The $p$-values of the VaR tests of the predicted violations for $\alpha=.92,.94,.94,.96,.98$.   `Ob.' represents the observed number of violations and
 `Exp.' represents the expected number of violations.  \label{table:var-test-honeypot1}}
\begin{tabular}{lcccccc}
  \toprule
 $\alpha$ &Exp. &Ob. &LRuc &LRcc
   \\\midrule
 $.92$ &41 &41 &0.9948 &.7118    \\\midrule
 $.94$ & 31 &31 &0.9585 & 0.9944   \\  \midrule
  $.95$&26 &23 &0.5919 & 0.5828 \\  \midrule
    $.96$&21 &17& 0.4192& 0.2124   \\  \midrule
   $.98$&10& 9 &0.6894 & 0.0241  \\
 \bottomrule
\end{tabular}
\end{table}

We conclude that the proposed approach has an overall satisfactory prediction performance.

\subsection{Honeypot data II with 69 dimensions}
In this section, we study the other honeypot data  which was used in \cite{peng2018modeling,zhan2013characterizing}. The dataset was collected by a low-interaction honeypot during 4 November 2010 to 21 December 2010 with a total number of 1123 hours, which has 69 consecutive IP addresses. Each TCP flow initiated by a remote computer
and an unsuccessful TCP handshake are deemed as attacks \cite{zhan2013characterizing}. The data is split into three parts: the training with {$750$} observations, the validation with {$150$} observations, and the rest $223$ observations are used for the prediction evaluation. The data is log-transformed to reduce the skewness.

We employ Algorithm \ref{alg1} for training and validating the deep learning model, and Algorithm \ref{alg2} for the prediction.  The performance of point predictions are shown in Table \ref{table:MAPE-mse-honeypot}.
\begin{table}[!htbp] \centering
  \caption{Summary statistics of predictions of proposed approach and benchmark model for MAPE and MSE based on the log-transformed attack data.\label{table:MAPE-mse-honeypot}}
 \begin{tabular}{@{\extracolsep{5pt}}lcccccccc}
\\[-1.8ex]\hline
\hline \\[-1.8ex]
Statistic & \multicolumn{1}{c}{N} & \multicolumn{1}{c}{Mean} & \multicolumn{1}{c}{St. Dev.} & \multicolumn{1}{c}{Min} & \multicolumn{1}{c}{$Q_1$} & \multicolumn{1}{c}{Median} & \multicolumn{1}{c}{$Q_3$} & \multicolumn{1}{c}{Max} \\
\hline \\[-1.8ex]
\multicolumn{9}{c}{Deep} \\ \hline
MAPE & 69 & 0.091 & 0.017 & 0.036 & 0.084 & 0.088 & 0.095 & 0.142 \\
MSE & 69 & 0.316 & 0.180 & 0.114 & 0.197 & 0.253 & 0.324 & 1.028 \\
\hline \\[-1.8ex]
\multicolumn{9}{c}{Benchmark} \\ \hline
MAPE & 69 & 0.150  & 0.033  & 0.041  & 0.139  & 0.162  & 0.167  & 0.189    \\
MSE & 69 & 0.718  & 0.184  & 0.152  & 0.625  & 0.716  & 0.861  & 1.092   \\
\hline\hline \\[-1.8ex]
\end{tabular}
\end{table}

\begin{figure}[htbp!]
\centering
\subfigure[MAPE-Deep]{\includegraphics[width=0.45\textwidth]{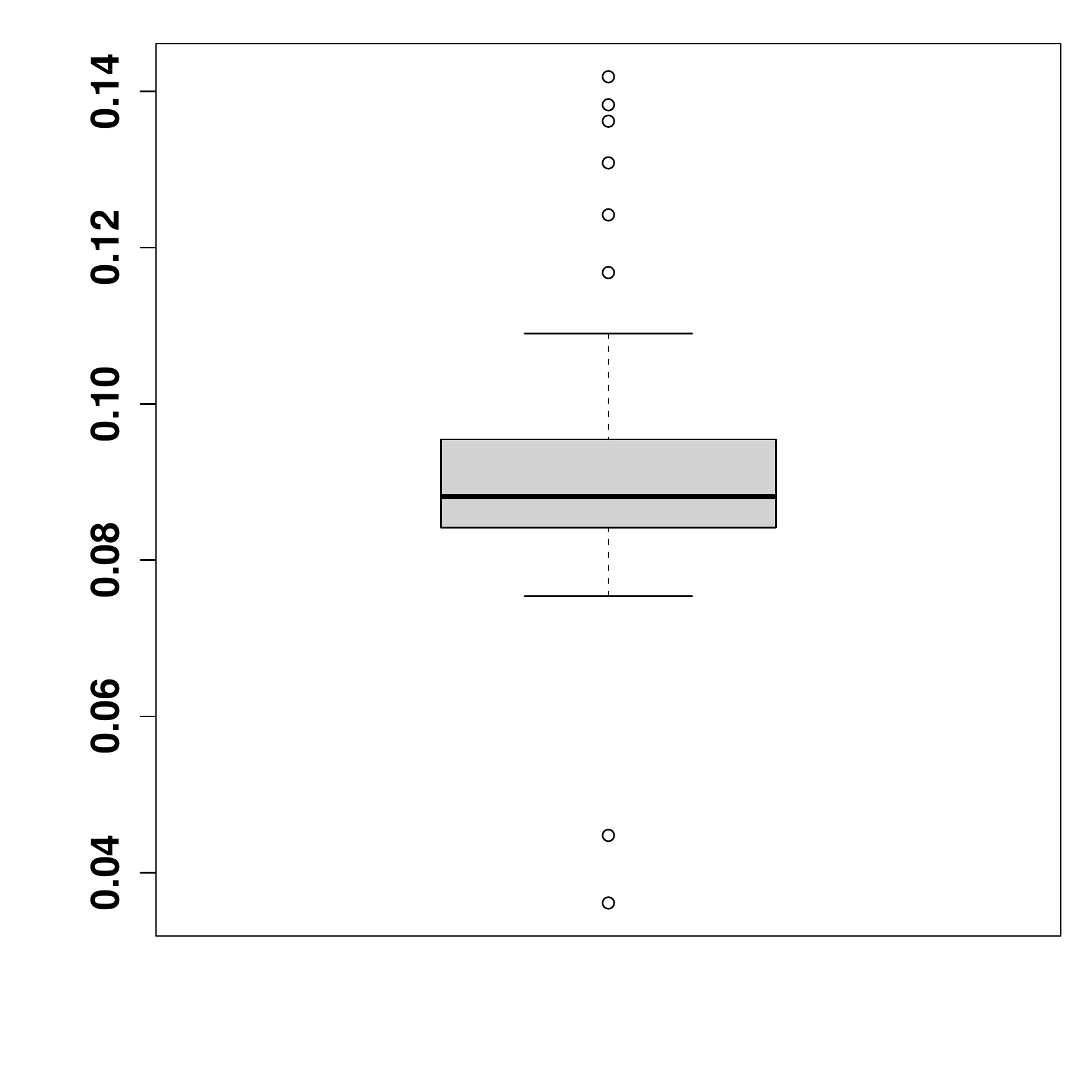}}
\subfigure[MAPE-Benchmark]{\includegraphics[width=0.45\textwidth]{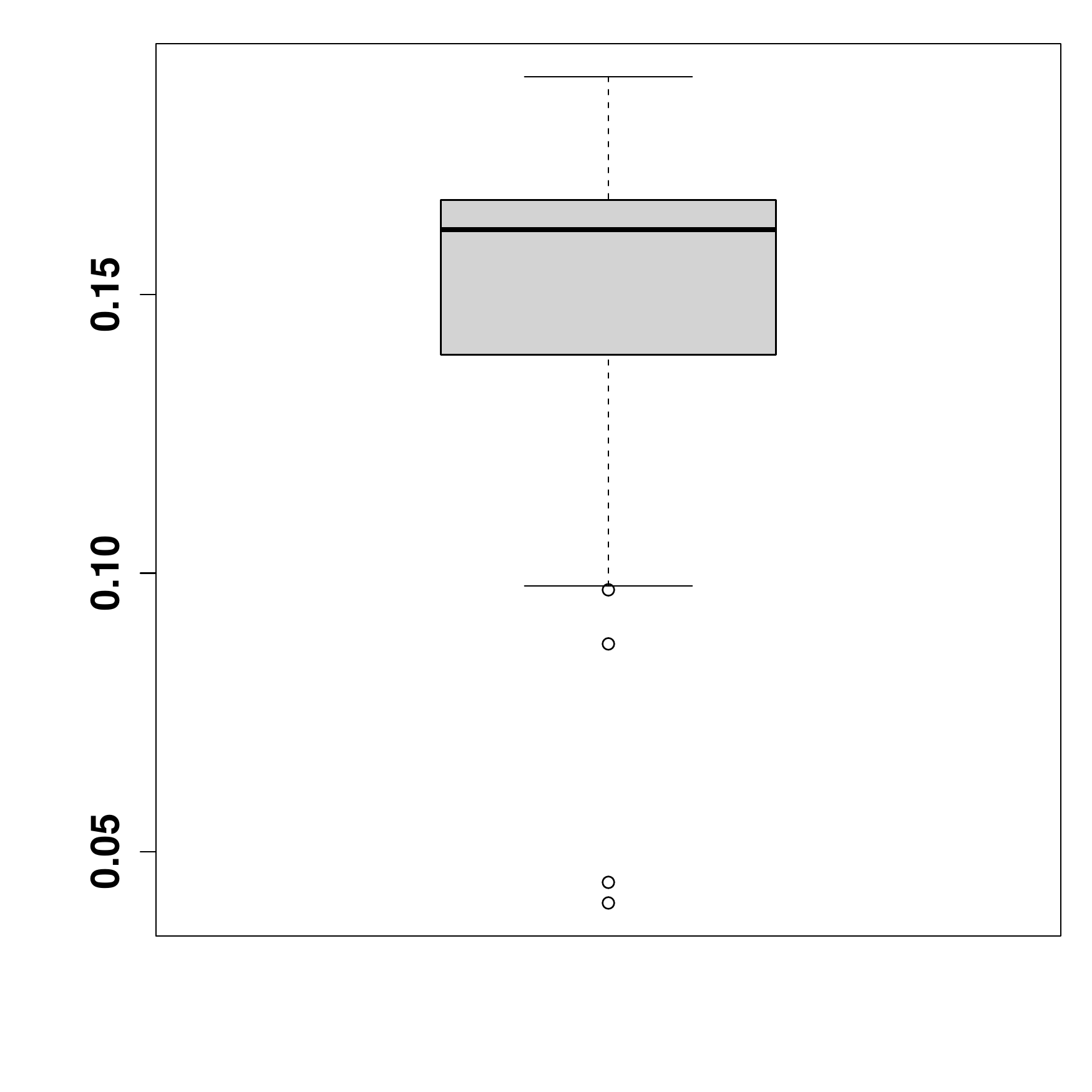}}
\subfigure[MSE-Deep]{\includegraphics[width=0.45\textwidth]{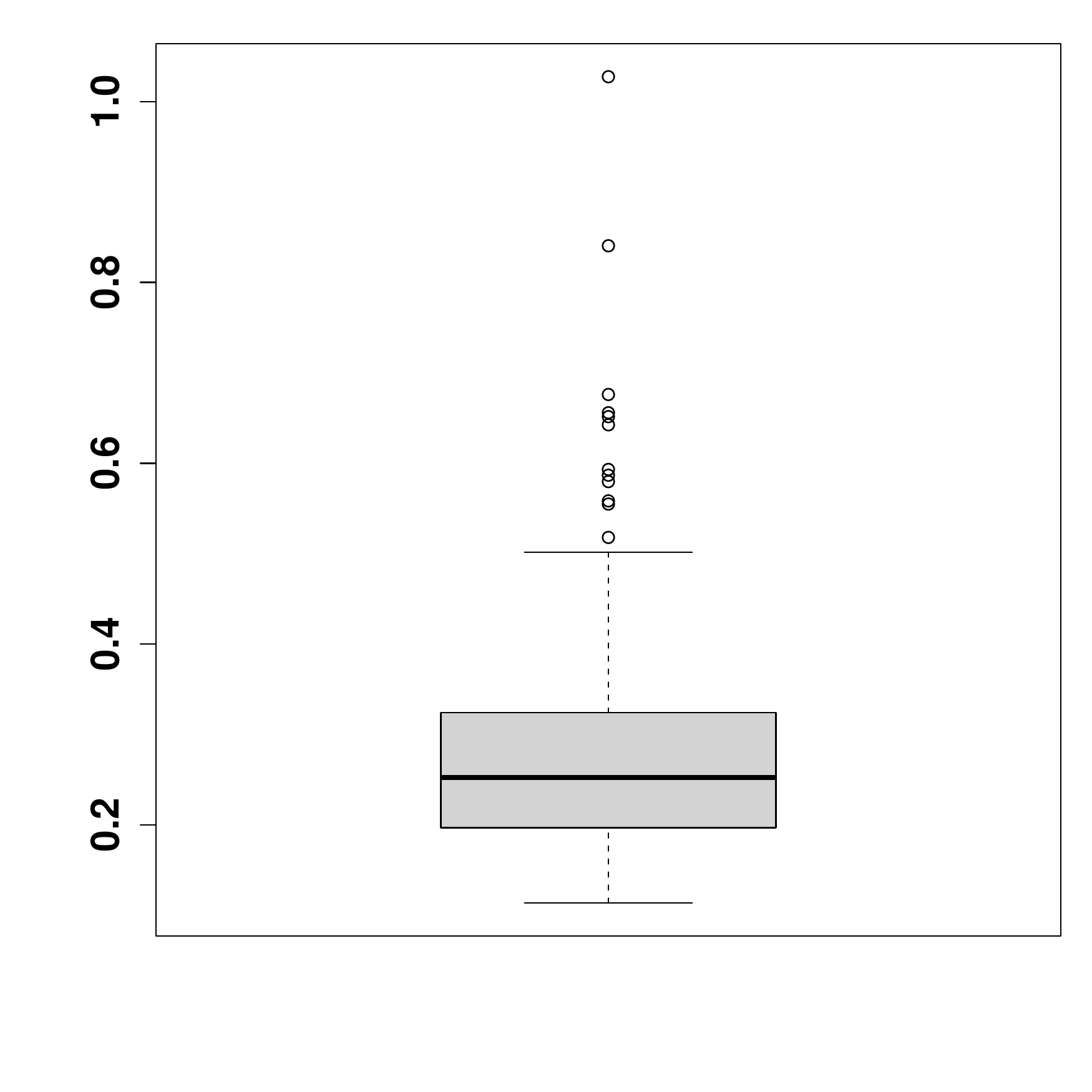}}
\subfigure[MSE-Benchmark]{\includegraphics[width=0.45\textwidth]{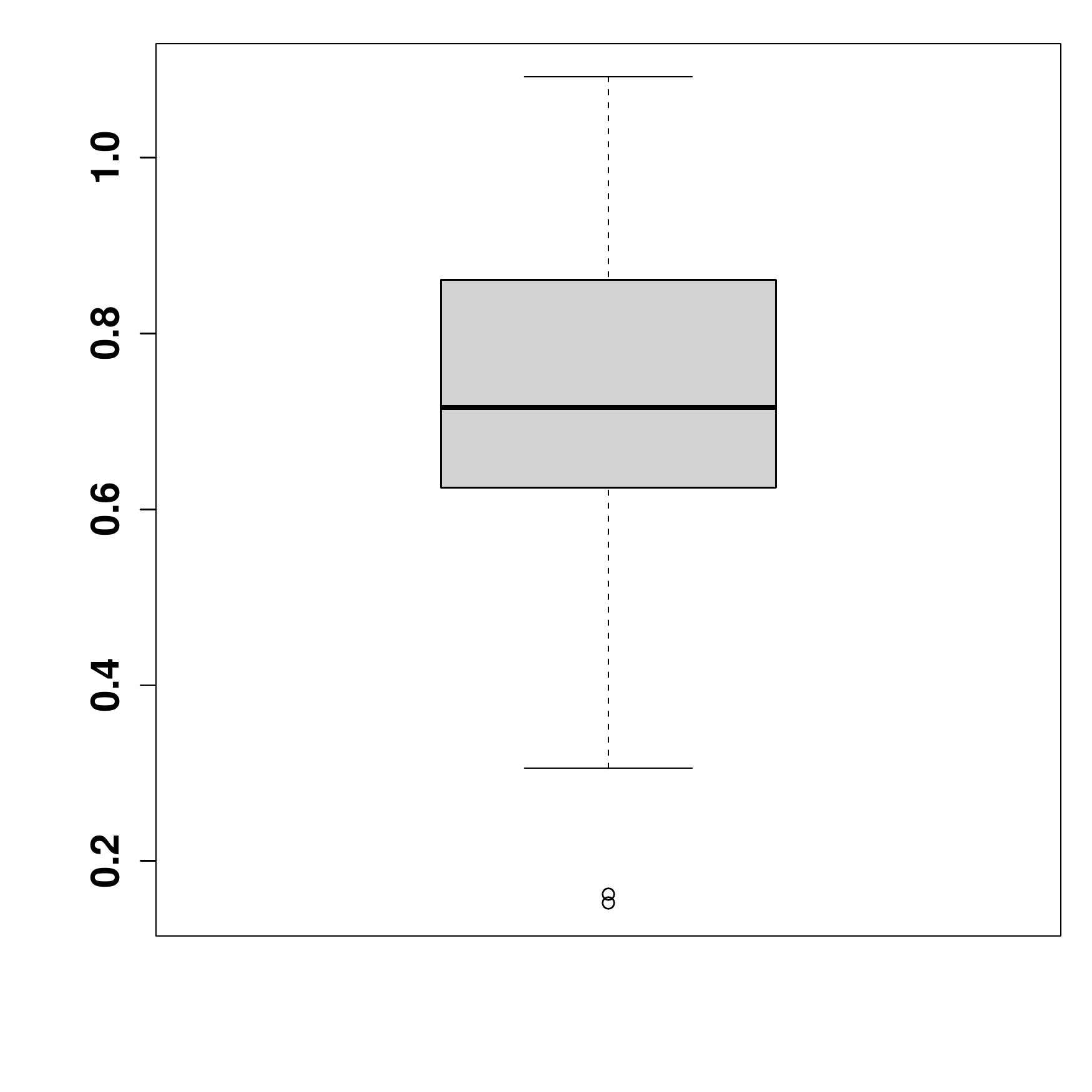}}
\caption{Boxplots of MAPEs and MSEs based on  the proposed model
and benchmark model for the log-transformed attack data.\label{fig:honepot-data2-boxplot-var}}
\end{figure}

It is observed that the proposed deep learning model has a very satisfactory prediction performance. Particularly, the mean (0.091) and median (0.088) of MAPEs by the proposed approach are much smaller than the mean (0.180) and median (0.162) of MAPEs by the benchmark model. For MSEs, the proposed approach also significantly outperforms the benchmark model in terms of mean (0.316 vs 0.718) and median (0.253 vs 0.716).
The boxplots are further shown in Figure
\ref{fig:honepot-data2-boxplot-var}. It is also seen that the proposed approach significantly outperforms the benchmark model.

\begin{figure}[htbp!]
\centering
\subfigure[QQ-plot]{\includegraphics[width=0.45\textwidth]{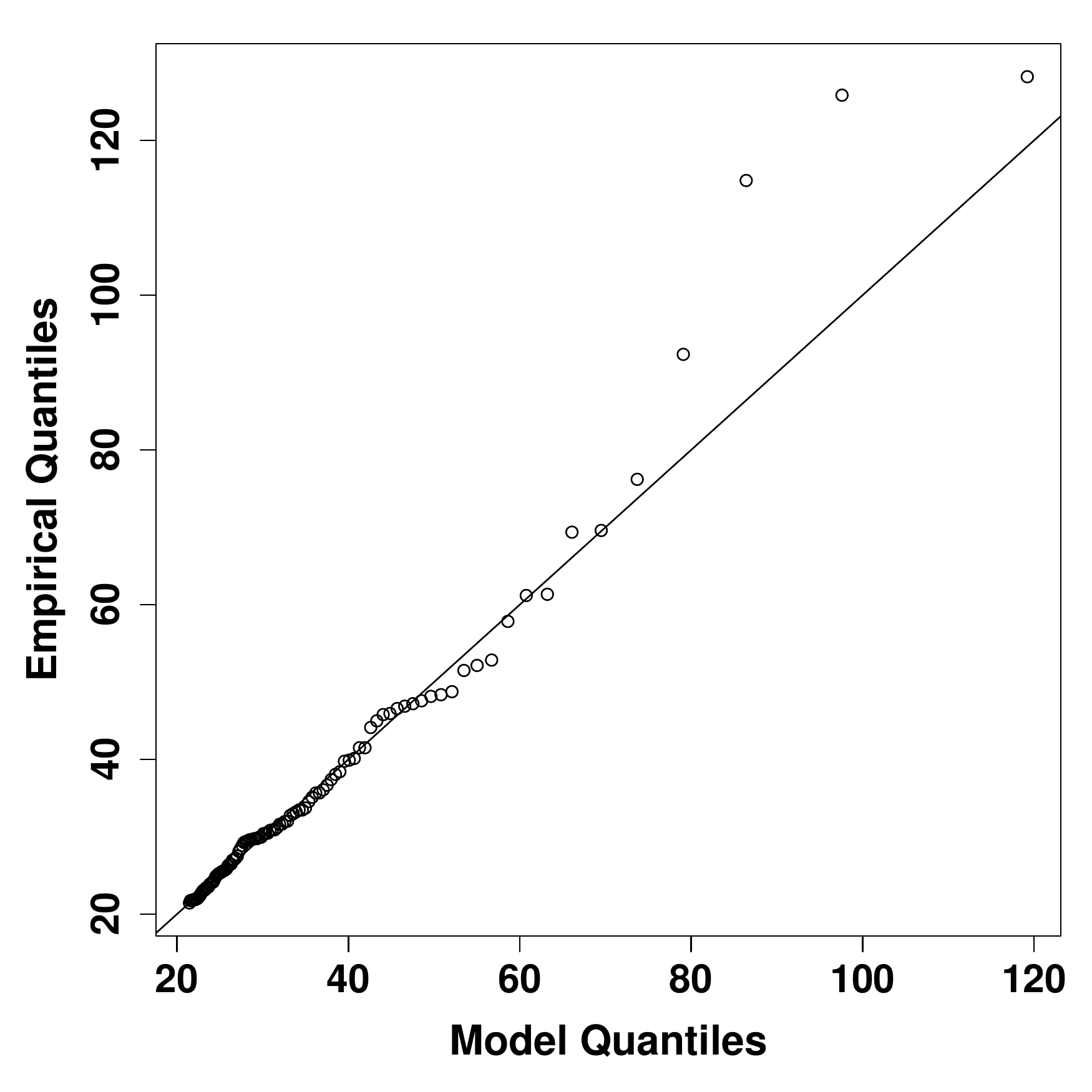}}
\subfigure[PP-plot]{\includegraphics[width=0.45\textwidth]{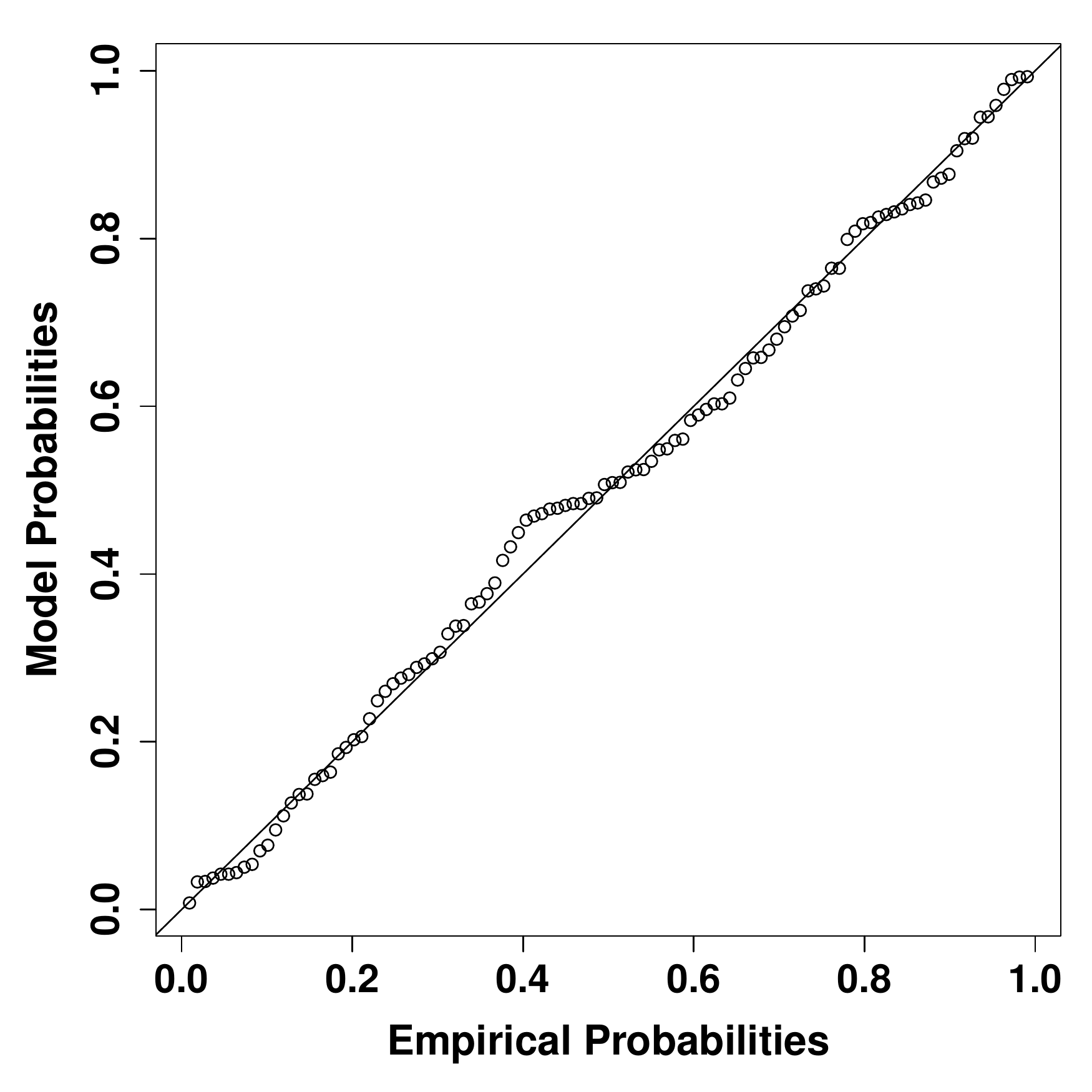}}
\caption{QQ-  and pp- plots of the log-transformed attack data fitted by the GPD distribution. \label{fig:honepot-data2-fngpd}}
\end{figure}
For the high quantiles, we also study the prediction performance at the aggregate level as it corresponds to the network level in practice where the defense can take place. Specifically, we calculate the residuals,
$$e_t=s_t - \hat s_t ,$$
where $s_t=\sum_{i=1}^{t}y'_i$,   $\hat s_t=\sum_{i=1}^{t} \hat y'_i$, and $y'$s are the log-transformed attack data,  $t=1,\ldots,900$. The GPD distribution is then fitted to $\{e_t\}_{t=1,\ldots,900}$ with the threshold to be $.9$-quantile of $\{e_t\}_{t=1,\ldots,900}$. The estimated parameters are $\xi=0.2116$ with standard error $0.1226$, and $\sigma=12.5542$ with standard error $2.0010$, which are significant at $.1$ level. The QQ- and PP- plots are displayed in Figure \ref{fig:honepot-data2-fngpd}. It is seen that PP- plot is satisfactory. The tails of QQ-plot is slightly off the diagonal line. This is again caused by the extreme large values in the attack data. But the overall fitting is fairly well.

For the high quantile prediction, Algorithm \ref{alg2} is employed to the testing data, where the threshold is set to be the .9-quantile of residuals. During the prediction process, we retrain the deep learning model three times by using Algorithm \ref{alg1} to further improve the prediction accuracy. The $p$-values of VaR tests are presented in Table \ref{table:honey69-test}. We first observe that the proposed model slightly underestimates the number of violations, particularly at the level $.95$ (11 vs 17), which is mainly due to the extreme large attacks. The other levels have large $p$-values which indicates that the proposed model can predict them well.
\begin{table}[htbp!]\centering
\centering
\caption{The $p$-values of the VaR tests of the predicted violations for $\alpha=.95,.96,.97,.98,.99$.   `Ob.' represents the observed number of violations and
 `Exp.' represents the expected number of violations.  \label{table:honey69-test}}
\begin{tabular}{lcccccc}
  \toprule
 $\alpha$ &Exp. &Ob. &LRuc &LRcc
   \\\midrule
 $.95$ &11 &17 &0.0846&0.0235  \\  \midrule
  $.96$&9 &12 &0.2960 &0.2022  \\  \midrule
   $.97$&7 &11 &0.1117 & 0.2392\\  \midrule
  $.98$& 4& 5& 0.7773& 0.8548   \\  \midrule
  $.99$& 2&4 &0.2738 &0.5100 \\
 \bottomrule
\end{tabular}
\end{table}

To conclude, the proposed model has  overall satisfactory fitting and prediction performances.
\section{Conclusion and discussion}\label{sec:conclusion}
In this work, we propose a novel approach to predicting the  multivariate cyber risks.  The key idea is to provide the accurate point prediction by training the deep learning network, while employ the extreme value theory for modeling and predicting the high quantiles. One particular advantage of proposed approach is that it can easily handle high dimensional cyber risks thanks to the deep learning architectures, and further, the deep learning model can be retrained to improve the prediction accuracy if needed.  The simulation and empirical studies confirm the feasibility of proposed approach and satisfactory fitting and prediction performances.

There are  some limitations for the current study. Since we model the multivariate dependence via deep learning architectures, the dependence is treated as a `black-box'. This causes the loss of interpreting the dependence relationship among cyber risks compared to the vine copula approach  \cite{peng2018modeling}, which is the common issue of deep learning approach. The real attack datasets studied in the current work were collected by honeypots. The other attack data (e.g. collected by telescope or other instruments) may exhibit different phenomena, which should be carefully analyzed before employing the approach developed in our work.

The current work can be extended in several directions.  For example, the deep learning model(s) may be further developed for more accurate fitting and prediction \cite{deng2014deep}. Second, for the complex multivariate cyber risks, the GPD distribution may not be enough for capturing the tail behavior. Then, the non-stationary extreme value distribution may be exploited for the possible modeling \cite{de2007extreme,mcneil2010quantitative}.

\begin{thebibliography}{10}

\bibitem{che2018recurrent}
Zhengping Che, Sanjay Purushotham, Kyunghyun Cho, David Sontag, and Yan Liu.
\newblock Recurrent neural networks for multivariate time series with missing
  values.
\newblock {\em Scientific reports}, 8(1):1--12, 2018.

\bibitem{cho2014learning}
Kyunghyun Cho, Bart Van~Merri{\"e}nboer, Caglar Gulcehre, Dzmitry Bahdanau,
  Fethi Bougares, Holger Schwenk, and Yoshua Bengio.
\newblock Learning phrase representations using rnn encoder-decoder for
  statistical machine translation.
\newblock {\em arXiv preprint arXiv:1406.1078}, 2014.

\bibitem{christoffersen1998evaluating}
Peter~F Christoffersen.
\newblock Evaluating interval forecasts.
\newblock {\em International economic review}, pages 841--862, 1998.

\bibitem{de2007extreme}
Laurens De~Haan and Ana Ferreira.
\newblock {\em Extreme value theory: an introduction}.
\newblock Springer Science \& Business Media, 2007.

\bibitem{deng2014deep}
Li~Deng, Dong Yu, et~al.
\newblock Deep learning: methods and applications.
\newblock {\em Foundations and Trends{\textregistered} in Signal Processing},
  7(3--4):197--387, 2014.

\bibitem{fang2019performance}
Xing Fang and Zhuoning Yuan.
\newblock Performance enhancing techniques for deep learning models in time
  series forecasting.
\newblock {\em Engineering Applications of Artificial Intelligence},
  85:533--542, 2019.

\bibitem{goel2017r2n2}
Hardik Goel, Igor Melnyk, and Arindam Banerjee.
\newblock R2n2: Residual recurrent neural networks for multivariate time series
  forecasting.
\newblock {\em arXiv preprint arXiv:1709.03159}, 2017.

\bibitem{Goodfellow-et-al-2016}
Ian Goodfellow, Yoshua Bengio, and Aaron Courville.
\newblock {\em Deep Learning}.
\newblock MIT Press, 2016.
\newblock \url{http://www.deeplearningbook.org}.

\bibitem{hochreiter1997long}
Sepp Hochreiter and J{\"u}rgen Schmidhuber.
\newblock Long short-term memory.
\newblock {\em Neural computation}, 9(8):1735--1780, 1997.

\bibitem{hyndman2006another}
Rob~J Hyndman and Anne~B Koehler.
\newblock Another look at measures of forecast accuracy.
\newblock {\em International journal of forecasting}, 22(4):679--688, 2006.

\bibitem{ponemon2020}
Ponemon Institute.
\newblock Cost of a data breach report.
\newblock
  \url{https://www.ibm.com/security/digital-assets/cost-data-breach-report/#/},
  September 2020.

\bibitem{JR2019}
Jay Jacobs and Bob Rudis.
\newblock Data-driven security: dataset collection.
\newblock
  \url{https://datadrivensecurity.info/blog/pages/dds-dataset-collection.html},
  September 2020.

\bibitem{jang2014survey}
Julian Jang-Jaccard and Surya Nepal.
\newblock A survey of emerging threats in cybersecurity.
\newblock {\em Journal of Computer and System Sciences}, 80(5):973--993, 2014.

\bibitem{kingma2014adam}
Diederik~P Kingma and Jimmy Ba.
\newblock Adam: A method for stochastic optimization.
\newblock {\em arXiv preprint arXiv:1412.6980}, 2014.

\bibitem{krause2016multiplicative}
Ben Krause, Liang Lu, Iain Murray, and Steve Renals.
\newblock Multiplicative lstm for sequence modelling.
\newblock {\em arXiv preprint arXiv:1609.07959}, 2016.

\bibitem{lai2018modeling}
Guokun Lai, Wei-Cheng Chang, Yiming Yang, and Hanxiao Liu.
\newblock Modeling long-and short-term temporal patterns with deep neural
  networks.
\newblock In {\em The 41st International ACM SIGIR Conference on Research \&
  Development in Information Retrieval}, pages 95--104, 2018.

\bibitem{langkvist2014review}
Martin L{\"a}ngkvist, Lars Karlsson, and Amy Loutfi.
\newblock A review of unsupervised feature learning and deep learning for
  time-series modeling.
\newblock {\em Pattern Recognition Letters}, 42:11--24, 2014.

\bibitem{ling2019predicting}
Xing Ling, Yeonwoo Rho, and Chee-Wooi Ten.
\newblock Predicting global trend of cybersecurity on continental honeynets
  using vector autoregression.
\newblock In {\em 2019 IEEE PES Innovative Smart Grid Technologies Europe
  (ISGT-Europe)}, pages 1--5. IEEE, 2019.

\bibitem{mcneil2000}
Alexander~J McNeil and R{\"u}diger Frey.
\newblock Estimation of tail-related risk measures for heteroscedastic
  financial time series: an extreme value approach.
\newblock {\em Journal of empirical finance}, 7(3):271--300, 2000.

\bibitem{mcneil2010quantitative}
Alexander~J McNeil, R{\"u}diger Frey, and Paul Embrechts.
\newblock {\em Quantitative risk management: concepts, techniques, and tools}.
\newblock Princeton university press, 2010.

\bibitem{DHS2015}
Department of~Homeland~Security.
\newblock National critical infrastructure security and resilience research and
  development plan.
\newblock
  \url{http://publish.illinois.edu/ciri-new-theme/files/2016/09/National-CISR-RD-Plan-Nov-2015.pdf},
  November 2015.

\bibitem{peng2018modeling}
Chen Peng, Maochao Xu, Shouhuai Xu, and Taizhong Hu.
\newblock Modeling multivariate cybersecurity risks.
\newblock {\em Journal of Applied Statistics}, 45(15):2718--2740, 2018.

\bibitem{sezer2020financial}
Omer~Berat Sezer, Mehmet~Ugur Gudelek, and Ahmet~Murat Ozbayoglu.
\newblock Financial time series forecasting with deep learning: A systematic
  literature review: 2005--2019.
\newblock {\em Applied Soft Computing}, 90:106181, 2020.

\bibitem{spitzner2003honeynet}
Lance Spitzner.
\newblock The honeynet project: Trapping the hackers.
\newblock {\em IEEE Security \& Privacy}, 1(2):15--23, 2003.

\bibitem{wang2020deeppipe}
Bin Wang, Tianrui Li, Zheng Yan, Guangquan Zhang, and Jie Lu.
\newblock Deeppipe: A distribution-free uncertainty quantification approach for
  time series forecasting.
\newblock {\em Neurocomputing}, 397(1):11--19, 2020.

\bibitem{xu2019cybersecurity}
Maochao Xu and Lei Hua.
\newblock Cybersecurity insurance: Modeling and pricing.
\newblock {\em North American Actuarial Journal}, 23(2):220--249, 2019.

\bibitem{xu2017vine}
Maochao Xu, Lei Hua, and Shouhuai Xu.
\newblock A vine copula model for predicting the effectiveness of cyber defense
  early-warning.
\newblock {\em Technometrics}, 59(4):508--520, 2017.

\bibitem{zhan2013characterizing}
Zhenxin Zhan, Maochao Xu, and Shouhuai Xu.
\newblock Characterizing honeypot-captured cyber attacks: Statistical framework
  and case study.
\newblock {\em IEEE Transactions on Information Forensics and Security},
  8(11):1775--1789, 2013.

\bibitem{zhan2015predicting}
Zhenxin Zhan, Maochao Xu, and Shouhuai Xu.
\newblock Predicting cyber attack rates with extreme values.
\newblock {\em IEEE Transactions on Information Forensics and Security},
  10(8):1666--1677, 2015.

\end{thebibliography}

\end{document}